\documentclass[aip,pop,floatfix,reprint]{revtex4-1}

\usepackage{graphicx,xcolor}
\usepackage{amssymb,amsmath}
\usepackage[normalem]{ulem}
\usepackage{upgreek}
\usepackage{cancel}

\usepackage{hyperref}
\hypersetup{
    colorlinks=true,
    citecolor=blue,
    linkcolor=blue,
    filecolor=blue,
    urlcolor=blue,
}

\newcommand{\rmb}{{\mathrm b}}

\newcommand{\rmc}{{\mathrm c}}
\newcommand{\rmd}{{\mathrm d}}

\newcommand{\rme}{{\mathrm e}}

\newcommand{\rmg}{{\mathrm g}}
\newcommand{\rmi}{{\mathrm i}}
\newcommand{\rmK}{{\mathrm K}}

\newcommand{\rmpp}{{\mathrm p}}
\newcommand{\rmq}{{\mathrm q}}

\begin{document}

\def\SaoPaulo{Instituto de Fisica, Universidade de Sao Paulo, 05508-090, S\~{a}o Paulo, S\~{a}o Paulo, Brazil}

\def\Marseille{Aix-Marseille Universit\'e, CNRS, UMR 7345 PIIM, 13397, Marseille, France}

\title{Wave-particle interactions in a long traveling wave tube with upgraded helix}

\author{M. C. de Sousa}
\email[]{meirielenso@gmail.com}
\affiliation\SaoPaulo
\affiliation\Marseille
\author{F. Doveil}
\email[]{fabrice.doveil@univ-amu.fr}
\affiliation\Marseille
\author{Y. Elskens}
\email[]{yves.elskens@univ-amu.fr}
\affiliation\Marseille
\author{I. L. Caldas}
\email[]{ibere@if.usp.br}
\affiliation\SaoPaulo

\begin{abstract}

We investigate the interaction of electromagnetic waves and electron beams in a 4 meters long traveling wave tube (TWT). The device is specially designed to simulate beam-plasma experiments without appreciable noise. This TWT presents an upgraded slow wave structure (SWS) that results in more precise measurements and makes new experiments possible. We introduce a theoretical model describing wave propagation through the SWS and validated by the experimental dispersion relation, impedance, phase and group velocities. We analyze nonlinear effects arising from the beam-wave interaction, such as the modulation of the electron beam and the wave growth and saturation process. When the beam current is low, the wave growth coefficient and saturation amplitude follow the linear theory predictions. However, for high values of current, nonlinear space charge effects become important and these parameters deviate from the linear predictions, tending to a constant value. After saturation, we also observe trapping of the beam electrons, which alters the wave amplitude along the TWT.

\end{abstract}

\maketitle

\section{Introduction}
\label{Sec:Introduction}

Wave-particle interactions are a nonlinear phenomenon \cite{Shukla1986, Elskens2003, Mendonca2001}, presenting regular and chaotic trajectories in phase space \cite{Escande1985, Escande1982, Lichtenberg1992}. Resonant islands can be used for particle acceleration \cite{Pakter1995, deSousa2010, deSousa2012, deSousa2018}, whereas chaotic orbits are responsible for particle heating \cite{Karney1978, CorreaSilva2013}. The linear regime for wave-particle interactions is well known, but many of its nonlinear aspects remain unclear.

This type of interaction is a fundamental process in plasmas \cite{Shukla1986, Elskens2003, Fisch1987, Berk1992}, particle beams and accelerators \cite{Shukla1986, Davidson2001, Edwards2004}. In particular, wave-particle interactions are the basis for electromagnetic radiation amplifiers, such as free electron lasers \cite{Shukla1986}, gyrotrons \cite{Gilmour2011}, traveling wave tubes \cite{Pierce1950, Gilmour1994, Gilmour2011} (TWTs), etc.

TWTs are vacuum electron devices \cite{Faillon2008, Gilmour2011} that present a broad bandwidth with a rather simple design. Industrial TWTs range from 2 to 30~cm in length, and are mainly used as signal amplifiers for wireless communications \cite{Minenna2019EPJH}, such as space telecommunication. On the other hand, longer TWTs (some meters long) can be used for basic plasma physics research \cite{Dimonte1977, Dimonte1978, Tsunoda1987, Tsunoda1991, Hartmann1995, Guyomarch1996, Doveil2005PRL, Doveil2005PPCF, MacorThesis2007, Doveil2011} since the equations that describe the TWT \cite{Pierce1950, Nordsieck1953, Tien1956, Gilmour1994} are the same as those characterizing the beam-plasma instability \cite{ONeil1971, ONeil1972} in the small cold beam limit \cite{Dimonte1977, Dimonte1978}.

Electromagnetic radiofrequency (rf) waves in the TWT propagate through a slow wave structure (SWS) and interact with an electron beam in a vacuum environment. Thus, it is possible to experimentally mimic a beam-plasma system without the effects caused by the background plasma, and we are able to properly identify the effects due to the beam dynamics. These characteristics make the TWT an extremely useful device to simulate one-dimensional beam-plasma systems, which represent a paradigm for instabilities in wave-particle interactions.

The first TWT used for plasma physics research was described by Dimonte and Malmberg \cite{Dimonte1977, Dimonte1978}. It was 3 meters long, built at the University of California in San Diego. The second research TWT \cite{Guyomarch1996, Doveil2005PRL, Doveil2005PPCF, Chandre2005, Doveil2006, MacorThesis2007, MacorEPJD2007, Doveil2011}, with 4 meters in length, was located at PIIM Laboratory, Aix-Marseille University (former Universit\'e de Provence). Both devices were helix TWTs \cite{Pierce1950, Gilmour1994, Gilmour2011}, with the helix supported by three alumina rods.

In this paper, we present the third TWT specially designed to simulate beam-plasma systems. This TWT is also located at PIIM Laboratory. It allows a great control of the waves and beam parameters, and contains a measurement system that provides information about both the waves and the beam. The TWT presents an upgraded SWS with the helix rigidly wrapped in a dielectric polyimide tape, which guarantees a more precise helix pitch along the whole device length. This reduces the uncertainty of the experimental data and allows us to work with arbitrary waveforms. Furthermore, the wave phase velocity is lower in the upgraded TWT. Resonant electrons also move slower and the interaction time between waves and particles is longer, resulting in the appearance of a great variety of nonlinear effects.

\begin{figure*}[!tb]
	\centering
	\includegraphics{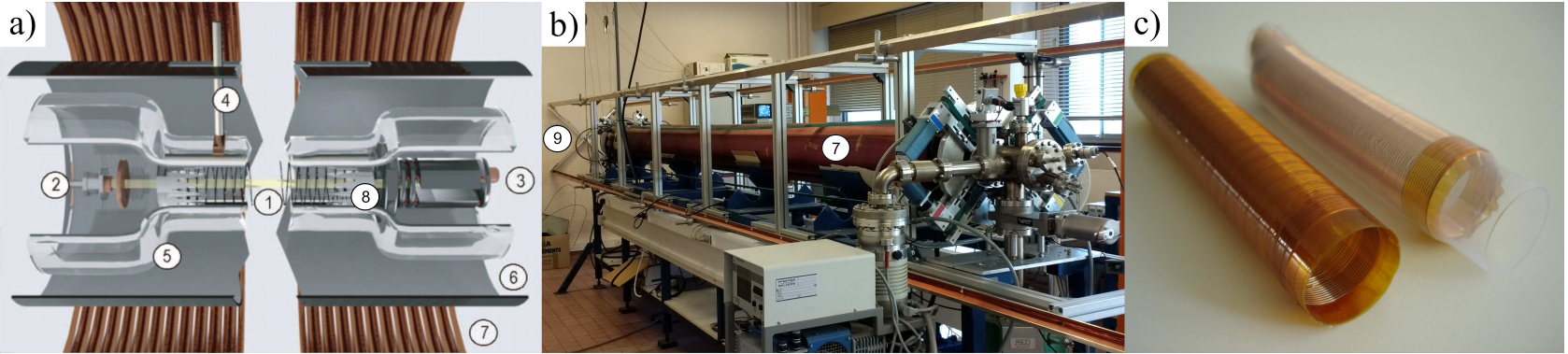}
	\caption{(Color online) (a) TWT structure (Reproduced from Ref. \onlinecite{Doveil2006}): (1) helix; (2) electron gun; (3) trochoidal velocity analyzer; (4) movable antenna; (5) glass vacuum tube; (6) slotted rf ground cylinder; (7) main magnetic coil; (8) resistive rf termination to reduce reflections. (b) Picture of the TWT showing its magnetic coils: (7) main coil producing the magnetic field $B_z$ to confine the beam, and (9) rectangular coils generating $B_x$ and $B_y$ for beam tilt correction. (c) Helix wrapped in a dielectric polyimide tape.}
	\label{fig:TWTStructure}
\end{figure*}

All these features of the upgraded TWT make new experiments possible, among which we may cite the use of pulsed beams \cite{MacorNPCS2007}, the experimental investigation of self-consistent effects \cite{Tennyson1994, Elskens2003, del-Castillo2002, Doveil2011}, and the quasilinear theory predictions \cite{Vedenov1962, Drummond1962, Vedenov1963, Tsunoda1987, Tsunoda1991, Hartmann1995, Elskens2007, Elskens2010, Elskens2010AAP, Besse2011, Elskens2012}. The upgraded TWT will also provide important experimental data for the validation of numerical codes \cite{Andre2013, Minenna2018, Minenna2019IEEE} that simulate wave-particle interactions in periodic structures. These new experiments and numerical simulations are important for plasma physics studies, but also contribute to the improvement of industrial devices.

We introduce a theoretical model describing the electromagnetic field through the upgraded SWS. We determine the dispersion relation, phase and group velocities, and we show that the theoretical parameters agree very well with the experimental data. We obtain experimentally the damping caused by the helix wire in the wave amplitude, and the voltage standing wave ratio (VSWR) that accounts for wave reflections inside the device. With these parameters, we completely characterize wave propagation in the upgraded TWT.

The interaction between waves and electrons is defined by the interaction impedance, or coupling impedance. We obtain the impedance both theoretically and experimentally with a very good agreement. The impedance decreases rapidly with the wave frequency, indicating a more efficient coupling for frequencies below 20~MHz.

We also investigate nonlinear effects occurring in the TWT. When the beam is emitted with initial velocity slightly higher than the wave phase velocity, electrons and wave enter in resonance. The wave receives momentum and energy from the beam, and its amplitude increases. This is the mechanism used by industrial TWTs to amplify telecommunication signals \cite{Minenna2019EPJH}. The TWT at PIIM Laboratory is 2 to 3 times longer than the length necessary for waves to saturate. After saturation, the beam electrons are trapped by the wave and form bunches that move back and forth in the wave potential, making the wave amplitude oscillate along the device.

We determine the wave growth coefficient and saturation amplitude. When the beam current is small, these parameters follow the predictions of the linear theory, proving that the wave saturates as a result of the development of electron bunches that are trapped in the wave potential. For higher values of current, we show that the growth coefficient and saturation amplitude deviate from the linear predictions due to nonlinear space charge effects caused by the repulsive electrostatic force among the beam electrons.

Another nonlinear effect analyzed in this paper is the modulation of the electron beam. An initially monokinetic beam gets modulated by the wave, and presents two distinct energy peaks at the end of the TWT. The difference between the two energy peaks provides a linear approximation, without damping effects, for the wave amplitude. We show that modulation occurs for electrons emitted with initial velocity both lower or higher than the wave phase velocity.

The paper is organized as follows. The experimental setup for the upgraded TWT is described in Section \ref{Sec:ExperimentalSetup}. In Section \ref{Sec:ColdParameters}, we develop the theoretical model for waves propagating in the TWT. We determine the theoretical and experimental dispersion relation, phase and group velocities, and we obtain experimentally the damping coefficient and VSWR. Section \ref{Sec:InteractionParameters} presents linear and nonlinear effects arising from the beam-wave interaction, including the modulation of the electron beam, the wave growth and saturation, the development of electrons bunches and the consequent oscillations in the wave amplitude. We calculate the four Pierce linear parameters \cite{Pierce1950, Gilmour1994, Guyomarch1996} that define the linear regime of TWTs. We show that the gain and space charge parameters increase with the beam current, meaning that nonlinear effects become important and the linear predictions lose accuracy for sufficiently high currents. In Section \ref{Sec:Conclusions}, we draw our conclusions and perspectives for the upgraded TWT.

\section{Experimental setup}
\label{Sec:ExperimentalSetup}

At PIIM Laboratory, we use a 4 meters long TWT specially conceived to study wave-particle interactions with applications in plasma physics. In the TWT, an electron beam moves in the axial direction, and it interacts with electromagnetic waves propagating through a helix waveguide. Near the axis, the magnetic field generated by the wave is negligible, and the electric field presents only longitudinal components, i.e.\ in the TWT axial direction. Therefore, electrons on the axis experience an electrostatic field as those observed in plasmas, which makes the TWT an ideal device to investigate wave-particle interactions in plasmas. Furthermore, the TWT at PIIM Laboratory is long enough for nonlinear effects to take place \cite{Chandre2005, MacorThesis2007}. The TWT can thus be used to mimic a one-dimensional beam-plasma experiment, with the advantages that it is much less noisy than any plasma and the medium supporting the waves is always in its linear regime.

The main components of the TWT are an electron gun, a trochoidal energy analyzer, and a SWS formed by a helix, where electromagnetic waves propagate \cite{Gilmour1994, Chandre2005}. In the TWT, it is possible to control several parameters with great accuracy. We use an arbitrary waveform generator that controls the number of modes produced, as well as the frequency, amplitude and phase of each individual mode. The electron beam is produced in such a way that we are able to determine its current, energy, and energy distribution function.

Figure~\ref{fig:TWTStructure} shows a schematic representation of the TWT at PIIM Laboratory. The most important part of the equipment is the SWS (labeled as (1) in Figure~\ref{fig:TWTStructure}(a)). It is composed of a 4 meters long helix made of a 0.5~mm diameter beryllium copper (BeCu) wire with a radius $a = 16.355$~mm. The helix has a small pitch $p = 1.0$~mm, so that waves traveling at the speed of light along the helix wire have a much smaller phase velocity along the axis. It guarantees that the waves interact resonantly with an electron beam propagating along the TWT axis.

In previous research TWTs \cite{Dimonte1977, Dimonte1978, Tsunoda1987, Tsunoda1991, Guyomarch1996, Doveil2005PRL, Doveil2005PPCF, Chandre2005, Doveil2006, MacorThesis2007, MacorEPJD2007, Doveil2011}, the helix was held inside a glass tube by three alumina rods. In this upgraded version of PIIM TWT, the helix is wrapped in and rigidly held by a dielectric polyimide tape (Figure~\ref{fig:TWTStructure}(c)), which ensures a nearly constant helix pitch along its full length, $0 \leq z \leq 4$~m. As we show along the paper, the results obtained with the upgraded helix are much more precise than those generated by the previous PIIM TWT. The experimental errors observed for the upgraded TWT are mainly due to the resolution of the equipments used for diagnostic and for launching the waves and the beam.

The helix is inserted into a glass vacuum tube with a resistive rf termination at each end to reduce wave reflections. The glass tube is evacuated by two ion pumps, one at each end of the device. The pressure inside the tube is typically on the order of $10^{-8}$~Torr. A good vacuum is necessary to avoid that the electron beam excites ions and forms a plasma in the system.

The glass vacuum tube is enclosed by an axially slotted cylinder with $R_5 = 57.5$~mm of radius that defines the rf ground. The TWT also contains four movable antennas capacitively coupled to the helix through the glass vacuum tube. Some of the antennas emit the waves produced by the arbitrary waveform generator. The other antennas move along the slotted cylinder to receive the spectrum of waves after interaction with the electron beam.

A triode (labeled as (2) in Figure~\ref{fig:TWTStructure}(a)) is located in one of the TWT extremities. It is used as an electron gun to produce a quasi-monoenergetic beam. The triode is composed of a heated cathode, a grid, and an anode with a small hole that determines the beam diameter (3~mm). The electron beam propagates along the axis of the SWS, and it is confined by an axial magnetic field $B_z$ generated by the main coil (Figure~\ref{fig:TWTStructure}(b)) that reaches a maximum value of 500~G. Two rectangular coils produce lower intensity magnetic fields, $B_x$ and $B_y$, on the order of 1~G for beam tilt correction.

A trochoidal energy analyzer \cite{Guyomarch2000} (labeled as (3) in Figure~\ref{fig:TWTStructure}(a)) is located in the other extremity of the TWT. The energy analyzer gives us the distribution function of energy in the beam with a resolution sharper than 0.5~eV. A small fraction ($\sim 0.5\%$) of the electrons passes through a hole in the center of the frontal collector, and it is decelerated by four electrodes. The electrons are then selected by the drift velocity caused by the presence of an electric field perpendicular to the magnetic field. Using this technique, it is possible to directly measure the current collected through a tiny off-axis hole, which gives us the time averaged axial energy distribution of the beam \cite{Guyomarch2000}.

\section{Wave propagation in the TWT}
\label{Sec:ColdParameters}

In this section, we analyze wave propagation in the TWT in absence of the electron beam. This propagation is characterized by the cold parameters: amplitude of the electromagnetic field generated by the waves through the SWS, dispersion relation, phase and group velocities, wave damping caused by the helix wire, and voltage standing wave ratio (VSWR) caused by wave reflections. We use Maxwell's equations to determine theoretically the electromagnetic field, dispersion relation, phase and group velocities. We compare the theoretical predictions with experimental data, and find an excellent agreement. The damping coefficient and the VSWR are obtained experimentally with great accuracy.

\subsection{Theoretical model}
\label{Sec:ColdParametersTheory}

A wave propagating at the speed of light $c$ along the helix wire has a much smaller velocity $v_z$ in the axial direction of the TWT. For a helix of radius $a$ and pitch $p$, we define $\tan\psi = p/(2 \pi a)$. The axial velocity $v_z$ may be approximated as $v_z = c \sin\psi$, which corresponds to $2.92 \times 10^6$~m/s for the upgraded TWT of PIIM Laboratory, with $\tan \psi = 0.00973 = 1/102.76$. The actual wave phase velocity $v_{\varphi}$ along the $z$ direction also depends on the other elements that compose the SWS, and is obtained through the dispersion relation calculated in this section.

The propagating wave generates electric and magnetic fields in the SWS given by Maxwell’s equations in Heaviside-Lorentz units \cite{Spohn2004}
\begin{gather} \label{eq:MaxwellEquations}
	\begin{aligned}
		\vec{\nabla} \cdot \vec{E} &= 0, \quad \quad & \vec{\nabla} \times \vec{E} = - \frac{1}{c} \frac{\partial \vec{B}}{\partial t} , \\*
		\vec{\nabla} \cdot \vec{B} &= 0, \quad \quad & \vec{\nabla} \times \vec{B} = \frac{\epsilon (r)}{c} \frac{\partial \vec{E}}{\partial t} ,
	\end{aligned}
\end{gather}
where $\epsilon (r)$ is the dielectric constant of the medium through which the electromagnetic wave propagates.

Considering a plane wave for which $\vec{E}, \, \vec{B} \, \sim \, \rme^{\rmi (kz - \omega t)}$, we calculate the components of the electromagnetic field in cylindrical coordinates and obtain the solution to equations (\ref{eq:MaxwellEquations}):
\begin{gather} \label{eq:EMField}
	\begin{aligned}
		E_{z,j} &= \left[ {C_j I_0 \left( {\gamma _j r} \right) + D_j K_0 \left( {\gamma _j r} \right) } \right] \rme^{\rmi (kz - \omega t)} , \\
		B_{z,j} &= \left[ {A_j I_0 \left( {\gamma _j r} \right) + B_j K_0 \left( {\gamma _j r} \right) } \right] \rme^{\rmi (kz - \omega t)} ,
	\end{aligned}
\end{gather}
where $I_0$ and $K_0$ are modified Bessel functions, $\omega$ is the angular frequency of the wave, $k$ is the (longitudinal) wavenumber, $\gamma _j$ is the (transversal) propagation constant of each medium given by
\begin{equation}
	\label{eq:Gamma}
		\gamma _j ^2 = k^2 - \epsilon _j \left( {\frac{\omega }{c}} \right) ^2 ,
\end{equation}
and the index $j$ indicates the medium through which the electromagnetic field propagates: $j=1$ vacuum (from the axis $r=0$~mm to the helix wire, which has an average radius $a = 16.355$~mm), $j=2$ dielectric tape (internal radius $R_1 = a$, external radius $R_2 = 16.515$~mm), $j=3$ vacuum ($R_2 < r < R_3$), $j=4$ glass vacuum tube (internal radius $R_3 = 17.1$~mm, external radius $R_4 = 22.25$~mm), $j=5$ air ($R_4 < r < R_5 = 57.5$~mm, with $R_5$ corresponding to the internal radius of the rf ground cylinder). In our model, we assume that all these structures are concentric.

We consider the helix as an infinitely thin, perfect conductor. It means that the electric field is null and the magnetic field is continuous inside the helix \cite{Jackson1999}:
\begin{gather} \label{eq:BoundaryConditions_r_a_E0}
	r = a \quad \Rightarrow \quad \left\{
	\begin{aligned}
		& E_{z,1} \sin\psi + E_{\theta ,1} \cos\psi = 0 , \\*
		& E_{z,2} \sin\psi + E_{\theta ,2} \cos\psi = 0 , \\*
		& B_{z,1} \sin\psi + B_{\theta ,1} \cos\psi \\*
			& \quad = B_{z,2} \sin\psi + B_{\theta ,2} \cos\psi .
	\end{aligned}
	\right.
\end{gather}
Moreover, the electric field components perpendicular to the radial direction are continuous \cite{Jackson1999}
\begin{gather} \label{eq:BoundaryConditions_r_a_EC}
	r = a \quad \Rightarrow \quad \left\{
	\begin{aligned}
		E_{z,1} &= E_{z,2} , \\*
		E_{\theta ,1} &= E_{\theta ,2} .
	\end{aligned}
	\right.
\end{gather}

The rf ground cylinder is also considered a perfect conductor. Therefore, we have
\begin{gather} \label{eq:BoundaryConditions_r_R5}
	r = R_5 \quad \Rightarrow \quad \left\{
	\begin{aligned}
		E_{z,5} &= 0 , \\*
		E_{\theta ,5} &= 0 .
	\end{aligned}
	\right.
\end{gather}

On the surface that separates two dielectric media, and in the case this surface does not contain localized electric charges or superficial currents, the components of the electric and magnetic fields are related as \cite{Jackson1999}
\begin{widetext}
	\begin{gather} \label{eq:BoundaryConditions_r_Rj}
		r = R_j \,(j = 2,3,4) \, \quad \Rightarrow \quad \left\{
		\begin{aligned}
			\epsilon _j E_{r,j} &= \epsilon _{j+1} E_{r, j+1} , \quad \quad & E_{\theta ,j} = E_{\theta ,j+1}, \quad \quad E_{z,j} &= E_{z,j+1} , \\*
			B_{r,j} &= B_{r, j+1}, \quad \quad & \frac{B_{\theta ,j}}{\mu _j} = \frac{B_{\theta , j+1}}{\mu _{j+1}}, \quad \quad \frac{B_{z,j}}{\mu _j} &= \frac{B_{z, j+1}}{\mu _{j+1}} ,
		\end{aligned}
		\right.
	\end{gather}
\end{widetext}
where $\mu _j$ is the magnetic permeability of medium $j$. We use the values $\epsilon _1 = \epsilon _3 = 1$ for the vacuum, $\epsilon _2 \cong 3.40$ for the dielectric polyimide tape at 1~MHz, $\epsilon _4 \cong 4.55$ for the Pyrex tube at 1~MHz, $\epsilon _5 \cong 1$ for the air, and $\mu _j \cong 1$ for all the dielectric materials in the SWS.

From equations (\ref{eq:EMField})-(\ref{eq:BoundaryConditions_r_Rj}), we calculate the coefficients of the electromagnetic field. For region $j=1$ that contains the helix axis, $B_1 = D_1 = 0$ to avoid divergences in (\ref{eq:EMField}). The first equation in (\ref{eq:BoundaryConditions_r_a_E0}) determines the ratio
\begin{equation} \label{eq:A1/C1}
	\frac{A_1}{C_1} = \rmi \frac{c \gamma_1}{\omega} \frac{I_0 (a \gamma_1)}{I_1 (a \gamma_1)} \tan \psi .
\end{equation}
From the other equations in (\ref{eq:BoundaryConditions_r_a_E0})-(\ref{eq:BoundaryConditions_r_Rj}), we obtain all the coefficients $A_j$, $B_j$, $C_j$ and $D_j$, with $2 \leq j \leq 5$, proportional to $C_1$.

\begin{figure}[!tb]
	\centering
	\includegraphics[width=1\linewidth]{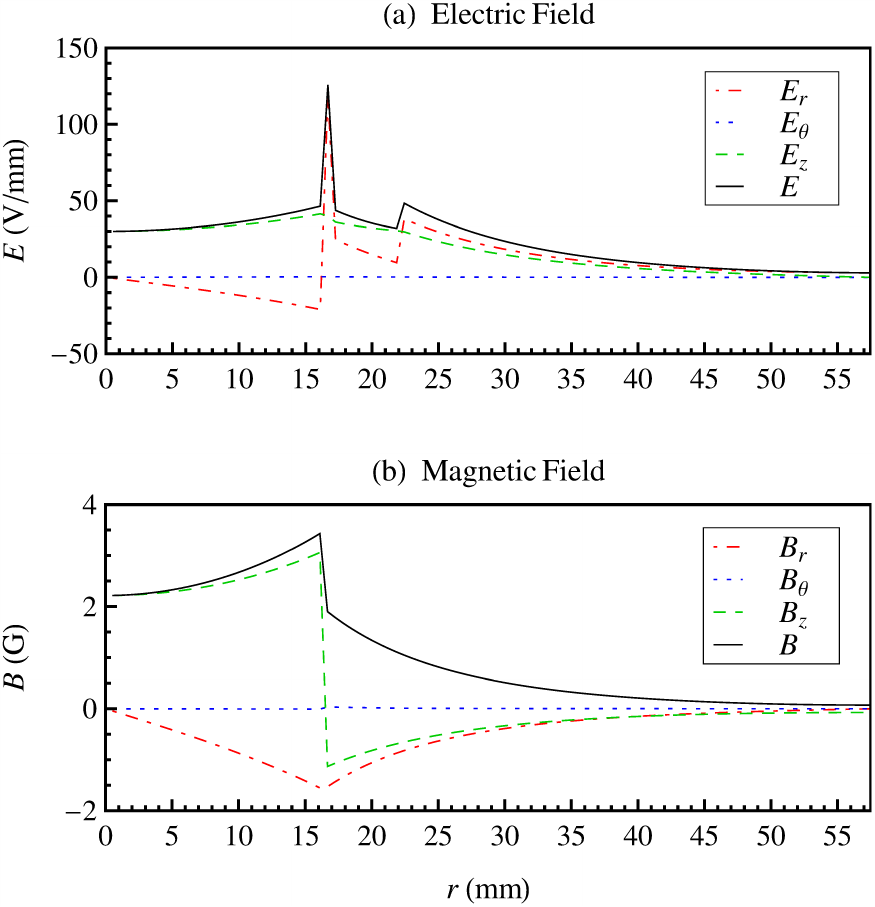}
	\caption{(Color online) Amplitude of each component of the (a) electric ($\vec{E}$) and (b) magnetic ($\vec{B}$) fields through a radial plane in the SWS. The wave that generates the field has a frequency 30~MHz.}
	\label{fig:EMField}
\end{figure}

Figure~\ref{fig:EMField} displays the amplitude of each component of the $\vec{E}$, $\vec{B}$ fields, with the normalization $C_1 = 1$~statV/cm $\cong 30$~V/mm, for a propagating wave with frequency 30~MHz. In the figure, it is possible to observe the behavior of the electromagnetic field through each individual component of the SWS. Near the axis ($r = 0$~mm), the electric and magnetic fields present only longitudinal components $E_z$, $B_z$. Electrons propagating along the TWT axis interact with an electrostatic field similar to the ones observed in plasmas.

The total amplitudes of the electric and magnetic fields reach their maximum value close to the helix ($r = a = 16.355$~mm). For the electric field, the radial component is the most important one near the helix. On the other hand, the radial and axial components of the magnetic field have comparable amplitudes. For both the electric and the magnetic field, the amplitude of the tangential component is null throughout the SWS. In the region near the rf ground cylinder ($r = R_5 = 57.5$~mm), all the fields components decay to zero. Numerical simulations show that the maximum value of the total electric and magnetic fields, $|\vec{E}| / C_1$ and $|\vec{B}| / C_1$, increases with the wave frequency. However, the qualitative behavior of the fields remains the same as in Figure~\ref{fig:EMField}.

The dispersion relation is obtained from the equation that ensures the continuity of the magnetic field at the helix wire:
\begin{equation} \label{eq:DispRelTransc}
	\left( {\dfrac{\omega }{c \gamma _2 \tan \psi }} \right) ^2 = \dfrac{\dfrac{Y_2 }{Y_1 } - \dfrac{\gamma _1 }{\gamma _2 } \dfrac{I_0 \left( {a \gamma _1 } \right) }{I_1 \left( {a \gamma _1 } \right) }}{\epsilon _2 \dfrac{Y_4 }{Y_3 } - \dfrac{\gamma _2 }{\gamma _1 } \dfrac{I_1 \left( {a \gamma _1 } \right) }{I_0 \left( {a \gamma _1 } \right) }} ,
\end{equation}
with $Y_j$ a function of the wave and SWS parameters. When the phase velocity $v_{\varphi} = \omega / k$ is much smaller than the speed of light, we may approximate
\begin{equation} \label{eq:GammaAppro}
	\gamma _j \approx k .
\end{equation}
Considering approximation (\ref{eq:GammaAppro}) in (\ref{eq:DispRelTransc}) yields
\begin{equation} \label{eq:DispRel}
	\omega = ck \tan \psi \left( {\dfrac{\dfrac{Y_2 }{Y_1 } - \dfrac{I_0 \left( {ak} \right) }{I_1 \left( {ak} \right) }}{\epsilon _2 \dfrac{Y_4 }{Y_3 } - \dfrac{I_1 \left( {ak} \right) }{I_0 \left( {ak} \right) }}} \right) ^{1/2} .
\end{equation}

From the dispersion relation (\ref{eq:DispRel}), we obtain the phase velocity $v_{\varphi }$ and the group velocity $v_\rmg$:
\begin{equation} \label{eq:vp_vg}
	v_{\varphi } = \frac{\omega }{k} , \quad \quad \quad v_\rmg = \frac{\partial \omega }{\partial k} .
\end{equation}	
\begin{figure}[!tb]
	\centering
	\includegraphics[width=1\linewidth]{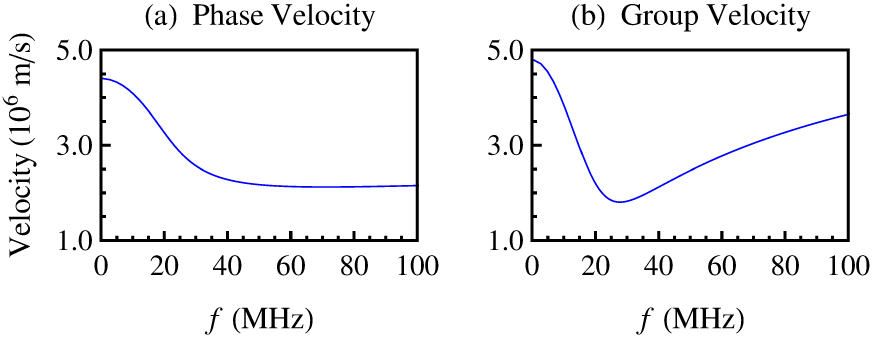}
	\caption{(a) Phase and (b) group velocities as a function of the wave frequency.}
	\label{fig:PhaseGroupVelocities}
\end{figure}
Figure~\ref{fig:PhaseGroupVelocities} shows the phase and group velocities as a function of the wave frequency $f$. The phase velocity decreases rapidly for frequencies between 0 and 40~MHz, and it is almost constant above 50~MHz. The group velocity also decreases rapidly for small frequencies. It presents a minimum around 27~MHz, and it increases again after this point.

\subsection{Experimental data}
\label{Sec:ColdParametersExperiments}

\begin{figure}[!tb]
	\centering
	\includegraphics[width=1\linewidth]{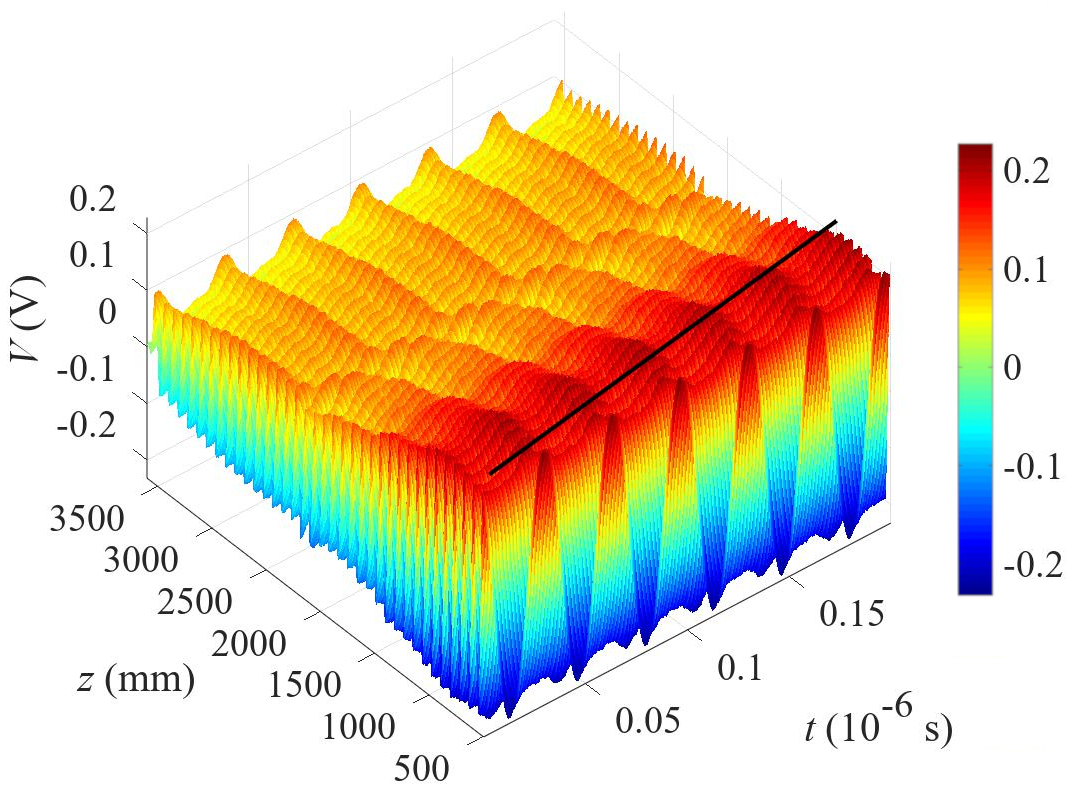}
	\caption{(Color online) Temporal signal registered by an oscilloscope for a propagating wave at 30~MHz. The black solid line indicates the theoretical phase velocity.}
	\label{fig:TemporalSignal}
\end{figure}

The antennas in the SWS are capacitively coupled to the helix through the glass vacuum tube. The signal is emitted by one of the antennas located near the electron gun. Another antenna moving axially along the TWT receives the signal that propagates along the helix wire. The temporal signal received by the moving antenna is registered by an oscilloscope and part of it is shown in Figure~\ref{fig:TemporalSignal} for a propagating wave at 30~MHz. The black line in Figure~\ref{fig:TemporalSignal} indicates the theoretical phase velocity $v_{\varphi }$ obtained from expression (\ref{eq:vp_vg}), and it agrees very well with the experimental data. The wave propagates all along the 4 meters TWT with the same phase velocity.

\begin{figure}[!tb]
	\centering
	\includegraphics[width=1\linewidth]{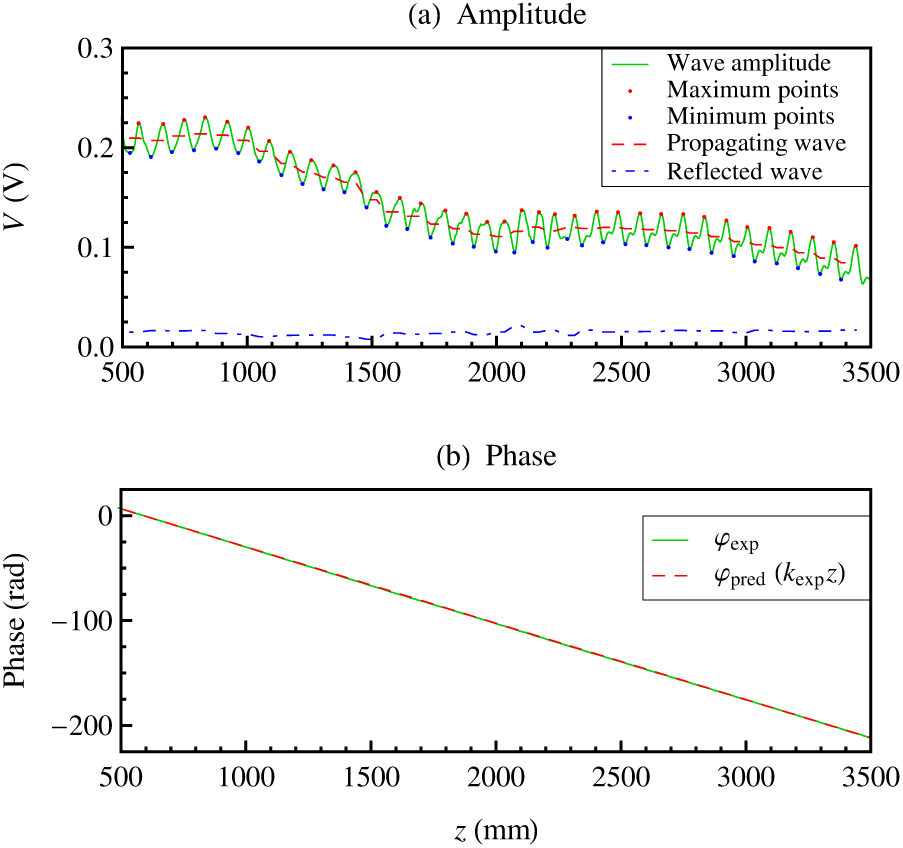}
	\caption{(Color online) (a) Experimental wave amplitude (green solid curve), and decomposition of total signal in two parts: propagating (red dashed curve) and reflected (blue dot-dashed curve) waves. (b) Experimental phase (green solid curve) and value predicted from the experimental wavenumber (red dashed curve). The wave amplitude and phase were obtained for the temporal signal in Figure~\ref{fig:TemporalSignal}.}
	\label{fig:DirRefWaveAmplitude}
\end{figure}

The temporal signal shown in Figure~\ref{fig:TemporalSignal} is registered by the oscilloscope for some given positions along the TWT (usually 900 -- 1800 different positions). For each position, we perform a Fast Fourier Transform (FFT) of the temporal signal to obtain its amplitude and phase. Figure~\ref{fig:DirRefWaveAmplitude} shows the experimental wave amplitude $V$ and phase $\varphi$ as a function of the axial position $z$ along the device for the temporal signal of Figure~\ref{fig:TemporalSignal}. In panel~(a), we notice the presence of 60~MHz harmonics for $z > 1600$~mm. Panel~(b) shows that the wave propagates along the 4 meters TWT with a uniformly varying phase, as predicted by the experimental wavenumber ($\varphi_{\mathrm{pred}} = k_{\mathrm{exp}}z$).

The experimental wavelength is obtained by interferometry. For each frequency, the interferometer multiplies the signal emitted by one of the antennas with the signal received by another antenna (as in Figure~\ref{fig:TemporalSignal}). We register the product of the signals as a function of the axial position of the receiver antenna. Through a numerical procedure, we determine the maxima of the registered signal, which gives us the averaged wavelength. The error is estimated as the standard deviation of the data points. The experimental wavelength can also be obtained through the FFT of the temporal signal. We determine the maximum points of the wave amplitude as shown in Figure~\ref{fig:DirRefWaveAmplitude}(a), and calculate the average wavelength. In both cases, the experimental wavelength agrees very well with the theoretical prediction (\ref{eq:DispRel}), with less than $1\%$ of difference.

\begin{figure*}[!tb]
	\centering
	\includegraphics[width=1\linewidth]{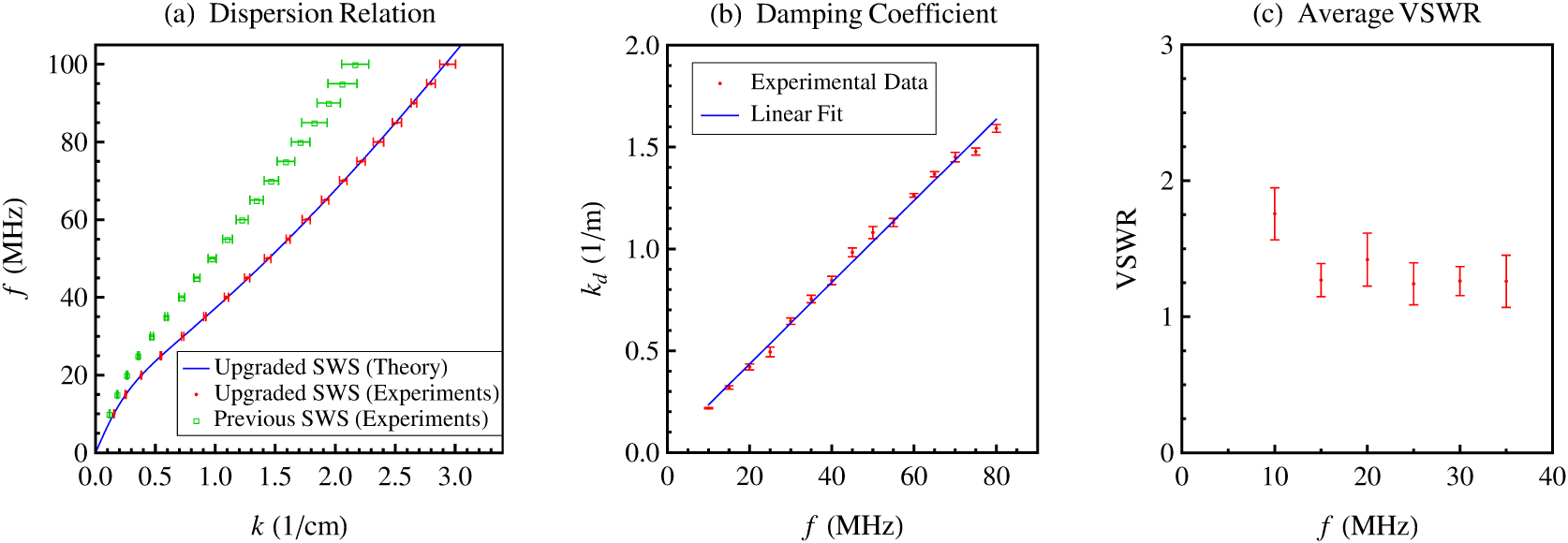}
	\caption{(Color online) (a) Theoretical dispersion relation (blue solid curve) and experimental data (full red circles) for the upgraded SWS, and experimental points (green open squares) for the previous SWS. (b) Damping coefficient and (c) average voltage standing wave ratio (VSWR) as a function of wave frequency for the upgraded SWS.}
	\label{fig:DispRelDampingStandingWave}
\end{figure*}

Figure~\ref{fig:DispRelDampingStandingWave}(a) shows the theoretical dispersion relation given by equation (\ref{eq:DispRel}) (blue solid curve) and the experimental points (full red circles) obtained by interferometry. The dispersion relation of the TWT closely resembles that of a finite radius, finite temperature plasma \cite{Malmberg1969}, but, unlike a plasma, the helix does not add appreciable noise.

Figure~\ref{fig:DispRelDampingStandingWave}(a) also shows the experimental points obtained for the previous version of the SWS (green open squares). By comparing the error bars for the two sets of experimental points, we observe that the upgraded helix and measurement system are much more precise than the previous one, which enables us not only to obtain more accurate experimental data, but also to carry out experiments that require a fine adjustment of the parameters. Furthermore, waves propagating along the upgraded SWS present a lower phase velocity. Electrons resonantly interacting with the wave move slower, resulting in a longer interaction time and the appearance of more nonlinear effects along the TWT.

In Figures \ref{fig:TemporalSignal} and \ref{fig:DirRefWaveAmplitude}(a), we observe a standing wave pattern, especially at the end of the device. Resistive rf terminations (labeled as (8) in Figure~\ref{fig:TWTStructure}(a)) are placed on both extremities of the glass tube to reduce wave reflections. However, residual reflections at the extremities and irregularities of the helix and glass tube generate a standing wave in the TWT.

The voltage standing wave ratio (VSWR) is defined as
\begin{equation} \label{eq:VSWR}
	\text{VSWR} = \frac{V_{\max{}}}{V_{\min{}}} ,
\end{equation}	
where $V_{\max{}}$ and $V_{\min{}}$ are, respectively, the local maximum and minimum values of the wave amplitude. In Figure~\ref{fig:DirRefWaveAmplitude}(a), we identify the maximum (red dots) and minimum (blue dots) points of the standing wave (green solid curve). With these points, we calculate the average VSWR along the TWT, as shown in Figure~\ref{fig:DispRelDampingStandingWave}(c). The error bars represent the standard deviation. For a propagating wave, i.e.\ no reflections, $\text{VSWR} = 1$. In the TWT, the VSWR varies between 1 and 2.

At the maximum points of the standing wave, the propagating and reflected waves are in phase and they interact constructively, so that $V_{\max{}} = V_{\mathrm{prop}} + V_{\mathrm{refl}}$. On the other hand, the waves are out of phase and they interact destructively at the minimum points: $V_{\min{}} = V_{\mathrm{prop}} - V_{\mathrm{refl}}$. Using this procedure, we decompose the total signal (green solid curve) in two parts representing the propagating (red dashed curve) and reflected (blue dot-dashed curve) waves, as can be seen in Figure~\ref{fig:DirRefWaveAmplitude}(a).

Figure~\ref{fig:DirRefWaveAmplitude}(a) shows that the wave is damped and its amplitude decreases along the TWT. In the absence of beam, the amplitude of the propagating wave varies as $V_{\mathrm{prop}} \sim \rme^{- k_\rmd z}$. Thus, from the propagating wave we obtain the experimental damping coefficient $k_\rmd$ of the helix for different wave frequencies, as shown in Figure~\ref{fig:DispRelDampingStandingWave}(b). By decomposing the signal and identifying the propagating wave, we determine the damping coefficient with great accuracy as shown by the small error bars in the figure. For the upgraded SWS, the damping coefficient is proportional to the wave frequency.

%

\section{Beam-wave interaction}
\label{Sec:InteractionParameters}

The interaction between waves and beam in the TWT is mainly characterized by the interaction impedance. We obtain the experimental impedance and show that it agrees with the theoretical predictions. The TWT at PIIM Laboratory is long enough to allow the appearance of nonlinear effects \cite{Chandre2005, MacorThesis2007}. In this section, we describe linear and nonlinear phenomena arising from the beam-wave interaction such as modulations in the electron distribution function, wave growth and saturation, and the development of electron bunches that alter the wave amplitude.

\subsection{Interaction impedance}
\label{Sec:Impedance}

The interaction impedance, also known as coupling impedance, characterizes the coupling between the electron beam and the wave electric field $E_z$ in the direction the beam propagates. The interaction impedance $Z_0$ is calculated theoretically as
\begin{equation} \label{eq:Z0}
	Z_0 = \frac{\left\langle {E_z^2} \right\rangle_\rmb }{2k^2 P} .
\end{equation}	
$\left\langle {E_z^2} \right\rangle_\rmb$ is the average value of $E_z^2$ over the transversal section $A_\rmb = \pi r_\rmb ^2$ of the electron beam with radius $r_\rmb = 1.5$~mm:
\begin{equation} \label{eq:Ez2}
	\left\langle {E_z^2} \right\rangle_\rmb = \frac{1}{A_\rmb } \int {E_z^2 \rmd A_\rmb } .
\end{equation}	
$P$ is the total wave power inside the rf ground cylinder given by
\begin{equation} \label{eq:Ptotal}
	P = \frac{1}{2} \text{Re} \left[ {\int {(\vec{S} \cdot \hat{z}) \rmd A_\rmc}} \right] ,
\end{equation}
with $\vec{S} = (\vec{E} \times \vec{B}) / \mu _0$ the Poynting vector, $\mu _0$ the permeability of free space, and $A_\rmc$ the transversal section of the rf ground cylinder.

\begin{figure}[!tb]
	\centering
	\includegraphics[width=0.55\linewidth]{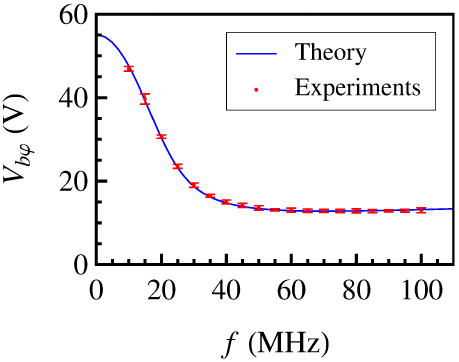}
	\caption{(Color online) $V_{\rmb \varphi}$ voltage for an electron beam with initial velocity $v_{\rmb 0}$ equal to the wave phase velocity $v_{\varphi}$.}
	\label{fig:BeamVoltage}
\end{figure}

To obtain the experimental interaction impedance, we need to define the voltage $V_{\rmb \varphi}$, which corresponds to the voltage applied to the electrons to create a beam with initial velocity $v_{\rmb 0}$ equal to the wave phase velocity $v_\varphi$. The beam voltage $V_{\rmb \varphi}$ is given by
\begin{equation} \label{eq:VbPhi}
	V_{\rmb \varphi} = \frac{m_\rme v_\varphi^2}{2e} ,
\end{equation}
where $m_\rme$ is the electron mass, and $e$ is the elementary charge. Figure~\ref{fig:BeamVoltage} presents the beam voltage $V_{\rmb \varphi}$ as a function of the wave frequency. The beam voltage decreases rapidly for frequencies below 40~MHz, and remains almost constant for higher frequencies. The blue solid curve was obtained from the theoretical dispersion relation using expressions (\ref{eq:DispRel}), (\ref{eq:vp_vg}) and (\ref{eq:VbPhi}). The red dots correspond to the experimental data in Figure~\ref{fig:DispRelDampingStandingWave}(a).

\begin{figure}[!tb]
	\centering
	\includegraphics[width=1\linewidth]{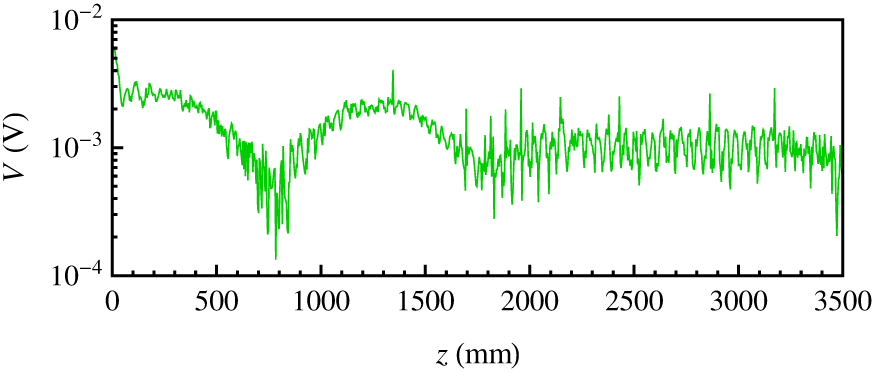}
	\caption{Wave amplitude (30~MHz, $V_{\rmb \varphi} = 19.1$~V) after interaction with an electron beam ($V_{\rmb 0} = 15$~V, $I_\rmb = 2$~$\upmu$A). The Kompfner dip is located at $z_{\rmK \rmd} = 783.1$~mm.}
	\label{fig:KompfnerDip}
\end{figure}

The original wave emitted by one of the TWT antennas produces a modulation in the electron beam, which in turn generates a second wave. The second wave induces another modulation in the beam, which produces a third wave. This process continues and generates a hierarchy of waves propagating in the TWT. However, when the wave has a small amplitude and the beam current $I_\rmb$ is low, the beam-wave interaction is well described by Pierce's three-wave model \cite{Pierce1950, Gilmour1994, Guyomarch1996}. In this case, and if the electrons initial velocity $v_{\rmb 0}$ is lower than the wave phase velocity $v_\varphi$, it is possible to find values of beam current and voltage for which the three waves interfere destructively in such a way that the total wave amplitude becomes null for a given position $z_{\rmK \rmd}$ (known as Kompfner dip) along the TWT axis. Figure~\ref{fig:KompfnerDip} shows a Kompfner dip observed for a wave emitted with $f = 30$~MHz, which corresponds to $V_{\rmb \varphi} \cong 19.1$~V. The electron beam was emitted with $V_{\rmb 0} = 15$~V and $I_\rmb = 2$~$\upmu$A. For this configuration, the total wave amplitude is null at $z_{\rmK \rmd} = 783.1$~mm.

\begin{figure}[!tb]
	\centering
	\includegraphics[width=0.63\linewidth]{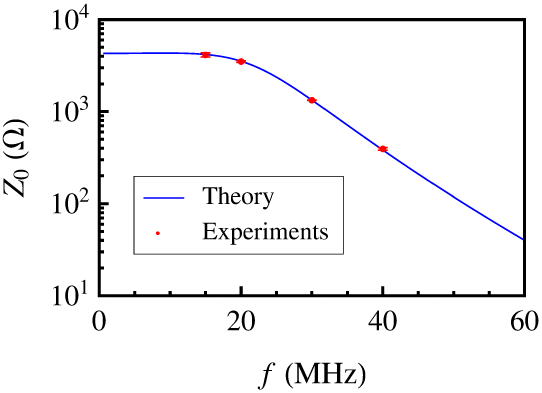}
	\caption{(Color online) Theoretical and experimental interaction impedance as a function of wave frequency. Error bars are within the marker size.}
	\label{fig:InteractionImpedance}
\end{figure}

The conditions to observe a null total wave amplitude were first described by Kompfner \cite{Kompfner1950}, and complemented later by Johnson \cite{Johnson1955}. We use the conditions and expressions described in these references, Pierce's three-wave model, and the parameters $V_{\rmb 0}$, $I_\rmb $, $z_{\rmK \rmd}$ obtained for the TWT to determine the experimental interaction impedance for the upgraded SWS. Figure~\ref{fig:InteractionImpedance} depicts the theoretical impedance (blue solid curve) calculated from expression (\ref{eq:Z0}), and the experimental values (red dots) obtained through the Kompfner dip method. Once again, theoretical and experimental values present a very good agreement. It shows the robustness of the theoretical model described in Section \ref{Sec:ColdParametersTheory}, and the accuracy of the experimental measurements for the upgraded version of the SWS and data acquisition system.

The electric and magnetic fields ($|\vec{E}|$, $|\vec{B}|$) present a peak near the helix, as can be seen in Figure~\ref{fig:EMField}. The peak value increases with the wave frequency, whereas the ($|\vec{E}|$, $|\vec{B}|$) values remain approximately constant near the TWT axis where the electron beam propagates. This means that the electromagnetic field gets more concentrated near the helix, and far from the beam, for higher frequencies, which results in a lower impedance. Figure~\ref{fig:InteractionImpedance} shows that the interaction impedance strongly decreases with the wave frequency, indicating that the coupling between particles and waves is less efficient for wave frequencies above 20~MHz.

%

\subsection{Electron velocity distribution and wave amplitude}
\label{Sec:VelocityDistribution}

%
\begin{figure}[!tb]
	\centering
	\includegraphics[width=1\linewidth]{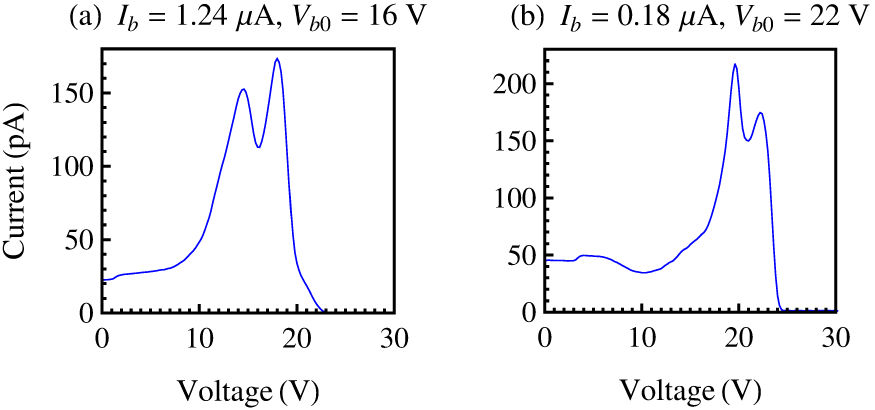}
	\caption{Electron distribution function showing the current collected by the trochoidal analyzer as a function of the electrons voltage at the end of the TWT. In panel~(a) the beam was emitted with $I_\rmb = 1.24$~$\upmu$A and $V_{\rmb 0} = 16$~V. For panel~(b) $I_\rmb = 0.18$~$\upmu$A and $V_{\rmb 0} = 22$~V. In both panels, the beam interacts with a 30~MHz wave ($V_{\rmb \varphi} = 19.1$~V) propagating through the SWS.}
	\label{fig:ElectronDistributionFunction}
\end{figure}

When waves interact with an electron beam, nonlinear effects take place such as the modulation of the beam. Figure~\ref{fig:ElectronDistributionFunction} shows the distribution function at the end of the TWT for two beams interacting with a 30~MHz wave ($V_{\rmb \varphi} = 19.1$~V). The electron gun generates a monokinetic beam with $V_{\rmb 0} = 16$~V in panel~(a), and $V_{\rmb 0} = 22$~V in panel~(b). After interacting with the wave along the device, the electron beams present distribution functions with peaks for two different values of voltage. For panel~(a), $V_{\rmb 0} = 16$~V, the peaks are centered around $V_{\rmb -} = 14$~V and $V_{\rmb +} = 18$~V, and the distribution function exhibits a local minimum for 16~V. This means that some electrons received energy from the wave reaching 18~V, while other electrons lost energy to the wave and were slowed down to 14~V. In panel~(b), the peaks of the distribution function are centered at $V_{\rmb +} = 22$~V, which is the initial beam voltage $V_{\rmb 0}$, and $V_{\rmb -} = 19$~V, corresponding to the voltage $V_{\rmb \varphi}$ of an electron beam propagating at the wave phase velocity for a 30~MHz wave.

The distribution functions in Figure~\ref{fig:ElectronDistributionFunction} are characteristic of beam modulation caused by its interaction with an electromagnetic wave. The difference between peaks in the distribution function can be used to estimate the wave amplitude $V_0$ disregarding the damping caused by the helix wire:
\begin{eqnarray} \label{eq:V0}
	V_0 & = & \frac{m_\rme}{2e} (v_{\rmb +} - v_{\rmb -}) |v_{\rmb 0} - v_{\varphi}|		\nonumber  \\*
	    & = & (\sqrt{V_{\rmb +}} - \sqrt{V_{\rmb -}}) |\sqrt{V_{\rmb 0}} - \sqrt{V_{\rmb \varphi}}| ,
\end{eqnarray}
where the velocities are obtained from $v_\rmb = \sqrt{2eV_\rmb /m_\rme}$. Using the linear approximation (\ref{eq:V0}), we estimate the wave amplitude as $V_0 = 0.18$~V for Figure~\ref{fig:ElectronDistributionFunction}(a), and $V_0 = 0.11$~V in Figure~\ref{fig:ElectronDistributionFunction}(b). Note that this amplitude is estimated directly from the wave effect on the beam, whereas the amplitudes recorded by the oscilloscope (in our other figures) are rescaled by the measurement chain.

\subsection{Wave growth and saturation}
\label{Sec:WaveGrowth}

%
\begin{figure}[!tb]
	\centering
	\includegraphics[width=1\linewidth]{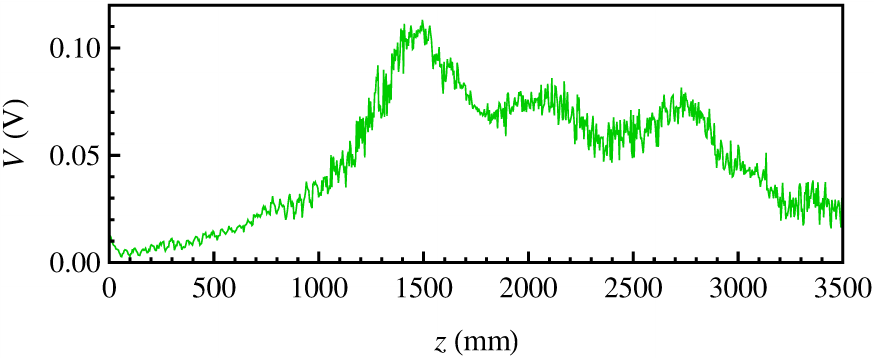}
	\caption{Wave amplitude (30~MHz) after interaction with an electron beam ($I_\rmb = 326$~$\upmu$A) emitted with higher velocity than the wave phase velocity.}
	\label{fig:ElectronPackets}
\end{figure}

The linear and nonlinear interaction between waves and particles can also produce the wave growth observed in Figure~\ref{fig:ElectronPackets}. Wave growth occurs for electron beams emitted with initial velocity $v_{\rmb 0}$ slightly higher than the wave phase velocity $v_\varphi$. In the beginning of the interaction process, the wave receives momentum and energy from the beam and its amplitude increases, as can be seen for $0 < z < 1500$~mm in Figure~\ref{fig:ElectronPackets}. This is the operation mechanism for industrial TWTs used as signal amplifiers \cite{Minenna2019EPJH}. The TWT at PIIM Laboratory is long enough for us to observe the development of electron bunches for $z > 1500$~mm, i.e.\ after the wave amplitude saturates. The electrons are trapped by the wave, moving back and forth in its potential. As a result of momentum and energy conservation, the wave amplitude oscillates along the TWT. The interaction between wave and electrons introduces noise in the signal, as shown in Figure~\ref{fig:ElectronPackets}, but the wave phase remains well defined.

To determine the wave growth coefficient, we begin by measuring the wave amplitude in absence of a beam (this is what we call signal 1). This signal presents effects related only to the SWS, such as the damping caused by the helix wire and the coupling between the helix and the receiving antenna. We then measure the wave amplitude in the presence of an electron beam (signal 2). Signal 2 contains effects related to both the SWS and the beam-wave interaction. By subtracting signal 1 from signal 2, we eliminate the influences caused by the SWS, and obtain a final signal that presents effects produced only by the beam-wave interaction.

In the growth region of final signal ($0 < z < 1500$~mm in Figure~\ref{fig:ElectronPackets} for example), the wave amplitude grows exponentially along the TWT as $V_{\mathrm{final}} \sim \rme^{k_\rmg z}$, with $k_\rmg$ the growth coefficient. Since the final signal contains only the effects caused by the beam-wave interaction, it enables us to determine the growth coefficient with great accuracy.

\begin{figure}[!tb]
	\centering
	\includegraphics[width=1\linewidth]{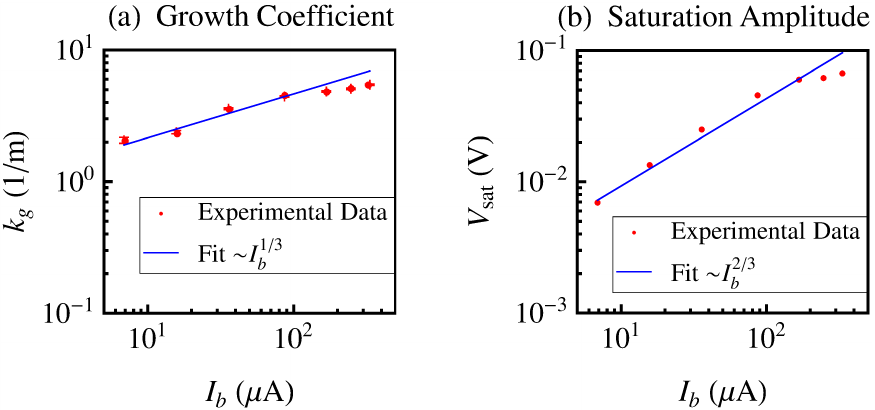}
	\caption{(Color online) (a) Growth coefficient and (b) saturation amplitude as a function of beam current for a wave emitted with 30~MHz.}
	\label{fig:GrowthSaturation}
\end{figure}

Figure~\ref{fig:GrowthSaturation}(a) displays the growth coefficient $k_\rmg$ as a function of the beam current $I_\rmb$ for a wave emitted at 30~MHz. The growth coefficient increases with the beam current, but it tends to a constant value for $I_\rmb \gtrsim 150$~$\upmu$A. The experimental data (red dots) for $I_\rmb < 150$~$\upmu$A agree with the theoretical prediction \cite{Pierce1950, Gilmour1994, Guyomarch1996} (blue solid curve), which estimates an increase in the growth coefficient proportional to $I_\rmb ^{1/3}$.

The saturation amplitude $V_{\mathrm{sat}}$ is the maximum amplitude reached by the wave at the end of the first growth stage. It is determined from signal 2, and corresponds to $z_{\mathrm{sat}} \sim 1500$~mm in Figure~\ref{fig:ElectronPackets}. The saturation amplitude varies with the beam current, as can be seen in Figure~\ref{fig:GrowthSaturation}(b). As well as the growth coefficient, the saturation amplitude increases for beam currents below 150~$\upmu$A, and tends to a constant value for $I_\rmb \gtrsim 150$~$\upmu$A.

When the wave saturates due to the development of electron bunches that are trapped by the wave potential, $V_{\mathrm{sat}}$ increases \cite{Dimonte1978, Guyomarch1996} with the beam current proportionally to $I_\rmb ^{2/3}$. The experimental data (red dots) in Figure~\ref{fig:GrowthSaturation}(b) agree very well with the theoretical prediction (blue solid curve), indicating that waves in the TWT saturate because of the nonlinear development of electron bunches along the device.

\subsection{Pierce linear parameters}
\label{Sec:PierceParams}

%
\begin{figure*}[!tb]
	\centering
	\includegraphics[width=1\linewidth]{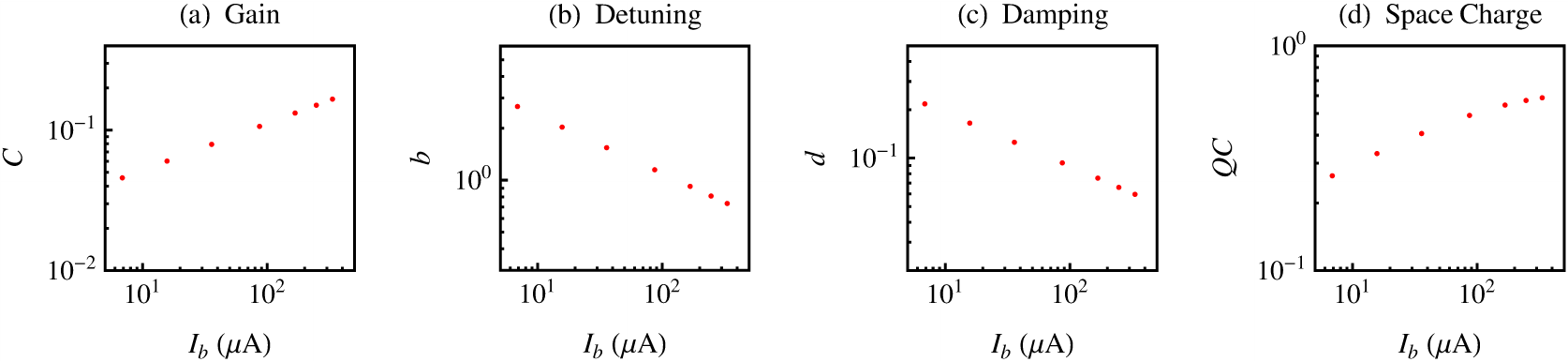}
	\caption{Pierce linear parameters: (a) gain, (b) detuning, (c) damping, and (d) space charge as a function of beam current for a 30~MHz wave and a beam emitted with $V_{\rmb 0} = 24$~V.}
	\label{fig:PierceParameters}
\end{figure*}

The linear regime of interaction between waves and beam in the TWT is completely characterized by four parameters \cite{Pierce1950, Gilmour1994, Guyomarch1996}, known as Pierce linear parameters. The gain parameter $C$ defines the wave gain as it interacts with the beam along the device:
\begin{equation} \label{eq:PierceParametersGain}
	C^3 = \frac{I_\rmb Z_0}{4V_{\rmb 0}} .
\end{equation}
The detuning parameter $b$ measures the normalized difference between the initial beam velocity and the wave phase velocity in the absence of electrons:
\begin{equation} \label{eq:PierceParametersDetuning}
	b = \frac{v_{\rmb 0} - v_{\varphi}}{C v_{\varphi}} .
\end{equation}
The damping parameter $d$ is the damping rate of the SWS in the absence of electrons normalized with the wave frequency, initial beam velocity and gain parameter:
\begin{equation} \label{eq:PierceParametersDamping}
	d = \dfrac{ k_\rmd }{C \omega / v_{\rmb 0}} .
\end{equation}
The space charge parameter $QC$ accounts for the repulsive electrostatic force between the beam electrons. It also takes into account the TWT geometry. Birdsall and Brewer \cite{Birdsall1954} calculated $QC$ as
\begin{equation} \label{eq:PierceParametersSpaceCharge}
	QC = \dfrac{1}{4C^2} \left( {\dfrac{{\omega _\rmq} / \omega}{1 + {\omega _\rmq} / \omega}} \right) ^2 ,
\end{equation}
where $\omega _\rmq = P_\rmq \omega _{\rmpp \rmb}$, with $\omega _{\rmpp \rmb} = (1/r_\rmb ) \sqrt{e I_\rmb / (\pi \epsilon_0 m_\rme v_{\rmb 0})}$ the beam plasma frequency, $\epsilon_0$ the vacuum permittivity, $P_\rmq = \left( {1 + R_\rmq ^2} \right) ^{- 1/2}$ the plasma frequency reduction factor due to the finite geometry of the beam \cite{Branch1955, Guyomarch1996}, $R_\rmq = v_{\rmb 0} \tau / (\omega r_\rmb) $, and $\tau$ a geometric factor of unitary order that varies slowly as a function of $\omega r_\rmb / v_{\rmb 0}$.

Figure~\ref{fig:PierceParameters} shows Pierce's linear parameters obtained from expressions (\ref{eq:PierceParametersGain})-(\ref{eq:PierceParametersSpaceCharge}) as a function of beam current for a 30~MHz wave, a constant beam voltage $V_{\rmb 0} = 24$~V, and $\tau = 1.3\, \omega r_\rmb / v_{\rmb 0} + 0.7228$ for the TWT. As expected, the gain parameter increases with the beam current. It means the wave extracts more energy and momentum from the beam, resulting in a higher growth coefficient and saturation amplitude as shown in Figure~\ref{fig:GrowthSaturation}. The detuning and damping parameters, on the other hand, decrease with the current.

In panel~\ref{fig:PierceParameters}(d), we observe that the space charge parameter increases with the beam current. For sufficiently high values of current, the electrostatic force acting on the beam electrons increases, the nonlinear effects caused by the beam space charge become important and the predictions of the linear theory lose accuracy. This is the case for the growth coefficient and saturation amplitude in Figure~\ref{fig:GrowthSaturation}, which deviate from the theoretical prediction for currents above 150~$\upmu$A.

The linear theory is valid for small enough values of the gain parameter, i.e.\ $C \ll 1$. For the upgraded TWT, we can estimate the beam current threshold for which the linear theory loses accuracy by considering $0.1<C<0.2$ in expression (\ref{eq:PierceParametersGain}), with $V_{\rmb 0}$ slightly higher than $V_{\rmb \varphi}$. Experimentally, we observe that the growth coefficient and saturation amplitude deviate from the linear predictions and tend to a constant value for $I_\rmb \gtrsim 150$~$\upmu$A for different values of wave frequency.

\section{Conclusions}
\label{Sec:Conclusions}

We analyzed the propagation of electromagnetic waves and electron beams, as well as their interaction, in an upgraded helix TWT. We presented a theoretical model describing the electromagnetic field through the SWS, and obtained the theoretical dispersion relation, phase and group velocities, and interaction impedance. We showed that the predicted theoretical parameters agree very well with the experimental data. It demonstrates the robustness of the model, as well as the good performance of the experimental device for its operating frequency range.

We also studied the nonlinear effects that take place in the TWT due to the beam-wave interaction. For an initially monokinetic beam, the distribution function gets modulated by the wave, presenting two peaks with different energies at the end of the device.

Another nonlinear effect occurs when the beam presents an initial velocity slightly higher than the wave phase velocity. In this case, the beam electrons resonantly interact with the wave. They transfer energy and momentum to the wave and its amplitude increases. After saturation, the wave amplitude oscillates along the TWT as the electrons form bunches that move back and forth in the wave potential.

We determined the wave growth coefficient and saturation amplitude as a function of the beam current. For sufficiently low values of current, we showed that these parameters increase with the current according to the linear prediction. Nonlinear effects are caused by the repulsive electrostatic force acting on the beam electrons. Such effects become important for high currents, and the growth coefficient and saturation amplitude deviate from the linear predictions, tending to a constant value.

The upgraded TWT of PIIM Laboratory presents a new configuration for the SWS. Usually, the helix in the SWS is held by three alumina rods. In the upgraded TWT, the helix is held by a dielectric polyimide tape rigidly wrapped all around the helix. It guarantees a more precise helix pitch along the 4 meters device, resulting in more accurate experimental measurements, and the possibility of working with different waveforms. Furthermore, waves propagating in the upgraded TWT present a lower phase velocity. This increases the interaction time for electrons resonantly interacting with the wave, and a variety of nonlinear effects can be observed.

All these features will allow us to perform new experiments to simulate wave-particle interactions in plasmas. Among the new experiments, we may cite the use of a pulsed beam \cite{MacorNPCS2007} instead of a continuous one, experiments to analyze the synergy between chaos and self-consistent effects \cite{Tennyson1994, Elskens2003, del-Castillo2002, Doveil2011}, and experiments to study the effects produced by the magnetic fields in the beam dynamics. In this paper, we considered the interaction of waves with a cold beam, and showed that waves saturate due to trapping of the beam electrons in the wave potential. In previous TWTs, experiments with a warm beam revealed that saturation in this case is caused by chaotic diffusion of the electrons in the broad spectrum excited by the beam \cite{Tsunoda1987, Tsunoda1991, Hartmann1995}. In a future work, we will use the upgraded TWT to carry out experiments with warm beams to investigate the predictions of the quasilinear theory \cite{Vedenov1962, Drummond1962, Vedenov1963, Tsunoda1987, Tsunoda1991, Hartmann1995, Elskens2007, Elskens2010, Elskens2010AAP, Besse2011, Elskens2012}.

Finally, this TWT may also be used to benchmak numerical models. Electromagnetic PIC (particle-in-cell) codes used for TWT simulations are usually too slow because of the great number of degrees of freedom to be considered. For this reason, PIC codes are not suitable for industrial applications that require faster simulations. As an alternative, a new time-domain code has been developed: DIMOHA \cite{Andre2013, Minenna2018, Minenna2019IEEE} (DIscrete MOdel with Hamiltonian Approach). The new code combines a Hamiltonian approach, that guarantees the respect of conservation laws, and an $N$-body description with a drastic reduction in the number of degrees of freedom. These characteristics allow DIMOHA to simulate nonlinear effects in TWTs much faster than traditional PIC codes \cite{Minenna2019IEEE}, enabling its use for industrial applications. DIMOHA simulates wave-particle interactions in periodic structures, and it has already been validated against industrial helix and folded waveguide TWTs \cite{Minenna2019IEEE} ($2-15$~cm long) and against the frequency-domain equivalent circuit Pierce model \cite{Minenna2019PhysScr}. The code will be upgraded to simulate long devices used for research in plasma physics, such as the 4 meters device at PIIM Laboratory. The experimental data obtained with the upgraded TWT will be used to validate the numerical results.

\begin{acknowledgments}

We thank D.~Guyomarc'h, J.-P.~Busso, J.-B.~Faure, V.~Long and J.-F.~Pioche for technical support with the experimental device, and Thales for contributing to the device upgrade.
We acknowledge financial support from the scientific agencies: S\~ao Paulo Research Foundation (FAPESP) under Grants No. 2013/01335-6, No. 2011/20794-6, No. 2015/05186-0 and No. 2018/03211-6, Coordena\c{c}\~ao de Aperfei\c{c}oamento de Pessoal de N{\'\i}vel Superior (CAPES) under Grants No. 88887.307684/2018-00 and No. 88881.143103/2017-01, and Comit\'e Fran\c{c}ais d'\'Evaluation de la Coop\'eration Universitaire et Scientifique avec le Br\'esil (COFECUB) under Grant No. 40273QA-Ph908/18.

\end{acknowledgments}

\section*{Data Availability}

The data that support the findings of this study are available from the corresponding authors upon reasonable request.

\bibliography{TWTBib}

\begin{thebibliography}{61}%
\makeatletter
\providecommand \@ifxundefined [1]{%
 \@ifx{#1\undefined}
}%
\providecommand \@ifnum [1]{%
 \ifnum #1\expandafter \@firstoftwo
 \else \expandafter \@secondoftwo
 \fi
}%
\providecommand \@ifx [1]{%
 \ifx #1\expandafter \@firstoftwo
 \else \expandafter \@secondoftwo
 \fi
}%
\providecommand \natexlab [1]{#1}%
\providecommand \enquote  [1]{``#1''}%
\providecommand \bibnamefont  [1]{#1}%
\providecommand \bibfnamefont [1]{#1}%
\providecommand \citenamefont [1]{#1}%
\providecommand \href@noop [0]{\@secondoftwo}%
\providecommand \href [0]{\begingroup \@sanitize@url \@href}%
\providecommand \@href[1]{\@@startlink{#1}\@@href}%
\providecommand \@@href[1]{\endgroup#1\@@endlink}%
\providecommand \@sanitize@url [0]{\catcode `\\12\catcode `\$12\catcode
  `\&12\catcode `\#12\catcode `\^12\catcode `\_12\catcode `\%12\relax}%
\providecommand \@@startlink[1]{}%
\providecommand \@@endlink[0]{}%
\providecommand \url  [0]{\begingroup\@sanitize@url \@url }%
\providecommand \@url [1]{\endgroup\@href {#1}{\urlprefix }}%
\providecommand \urlprefix  [0]{URL }%
\providecommand \Eprint [0]{\href }%
\providecommand \doibase [0]{http://dx.doi.org/}%
\providecommand \selectlanguage [0]{\@gobble}%
\providecommand \bibinfo  [0]{\@secondoftwo}%
\providecommand \bibfield  [0]{\@secondoftwo}%
\providecommand \translation [1]{[#1]}%
\providecommand \BibitemOpen [0]{}%
\providecommand \bibitemStop [0]{}%
\providecommand \bibitemNoStop [0]{.\EOS\space}%
\providecommand \EOS [0]{\spacefactor3000\relax}%
\providecommand \BibitemShut  [1]{\csname bibitem#1\endcsname}%
\let\auto@bib@innerbib\@empty
\bibitem [{\citenamefont {Shukla}\ \emph {et~al.}(1986)\citenamefont {Shukla},
  \citenamefont {Rao}, \citenamefont {Yu},\ and\ \citenamefont
  {Tsintsadze}}]{Shukla1986}%
  \BibitemOpen
  \bibfield  {author} {\bibinfo {author} {\bibfnamefont {P.~K.}\ \bibnamefont
  {Shukla}}, \bibinfo {author} {\bibfnamefont {N.~N.}\ \bibnamefont {Rao}},
  \bibinfo {author} {\bibfnamefont {M.~Y.}\ \bibnamefont {Yu}}, \ and\ \bibinfo
  {author} {\bibfnamefont {N.~L.}\ \bibnamefont {Tsintsadze}},\ }\href
  {\doibase 10.1016/0370-1573(86)90157-2} {\bibfield  {journal} {\bibinfo
  {journal} {Physics Reports}\ }\textbf {\bibinfo {volume} {138}},\ \bibinfo
  {pages} {1} (\bibinfo {year} {1986})}\BibitemShut {NoStop}%
\bibitem [{\citenamefont {Elskens}\ and\ \citenamefont
  {Escande}(2003)}]{Elskens2003}%
  \BibitemOpen
  \bibfield  {author} {\bibinfo {author} {\bibfnamefont {Y.}~\bibnamefont
  {Elskens}}\ and\ \bibinfo {author} {\bibfnamefont {D.~F.}\ \bibnamefont
  {Escande}},\ }\href@noop {} {\emph {\bibinfo {title} {Microscopic dynamics of
  plasmas and chaos}}}\ (\bibinfo  {publisher} {IOP Publishing},\ \bibinfo
  {address} {Bristol},\ \bibinfo {year} {2003})\BibitemShut {NoStop}%
\bibitem [{\citenamefont {Mendon{\c{c}}a}(2001)}]{Mendonca2001}%
  \BibitemOpen
  \bibfield  {author} {\bibinfo {author} {\bibfnamefont {J.~T.}\ \bibnamefont
  {Mendon{\c{c}}a}},\ }\href@noop {} {\emph {\bibinfo {title} {Theory of photon
  acceleration}}}\ (\bibinfo  {publisher} {IOP Publishing},\ \bibinfo {address}
  {Bristol},\ \bibinfo {year} {2001})\BibitemShut {NoStop}%
\bibitem [{\citenamefont {Escande}(1985)}]{Escande1985}%
  \BibitemOpen
  \bibfield  {author} {\bibinfo {author} {\bibfnamefont {D.~F.}\ \bibnamefont
  {Escande}},\ }\href {\doibase 10.1016/0370-1573(85)90019-5} {\bibfield
  {journal} {\bibinfo  {journal} {Physics Reports}\ }\textbf {\bibinfo {volume}
  {121}},\ \bibinfo {pages} {165} (\bibinfo {year} {1985})}\BibitemShut
  {NoStop}%
\bibitem [{\citenamefont {Escande}(1982)}]{Escande1982}%
  \BibitemOpen
  \bibfield  {author} {\bibinfo {author} {\bibfnamefont {D.~F.}\ \bibnamefont
  {Escande}},\ }\href {\doibase 10.1088/0031-8949/1982/t2a/016} {\bibfield
  {journal} {\bibinfo  {journal} {Physica Scripta}\ }\textbf {\bibinfo {volume}
  {T2/1}},\ \bibinfo {pages} {126} (\bibinfo {year} {1982})}\BibitemShut
  {NoStop}%
\bibitem [{\citenamefont {Lichtenberg}\ and\ \citenamefont
  {Lieberman}(1992)}]{Lichtenberg1992}%
  \BibitemOpen
  \bibfield  {author} {\bibinfo {author} {\bibfnamefont {A.~J.}\ \bibnamefont
  {Lichtenberg}}\ and\ \bibinfo {author} {\bibfnamefont {M.~A.}\ \bibnamefont
  {Lieberman}},\ }\href@noop {} {\emph {\bibinfo {title} {Regular and chaotic
  dynamics}}},\ \bibinfo {edition} {2nd}\ ed.\ (\bibinfo  {publisher}
  {Springer},\ \bibinfo {address} {New York},\ \bibinfo {year}
  {1992})\BibitemShut {NoStop}%
\bibitem [{\citenamefont {Pakter}\ and\ \citenamefont
  {Corso}(1995)}]{Pakter1995}%
  \BibitemOpen
  \bibfield  {author} {\bibinfo {author} {\bibfnamefont {R.}~\bibnamefont
  {Pakter}}\ and\ \bibinfo {author} {\bibfnamefont {G.}~\bibnamefont {Corso}},\
  }\href {\doibase 10.1063/1.870986} {\bibfield  {journal} {\bibinfo  {journal}
  {Physics of Plasmas}\ }\textbf {\bibinfo {volume} {2}},\ \bibinfo {pages}
  {4312} (\bibinfo {year} {1995})}\BibitemShut {NoStop}%
\bibitem [{\citenamefont {{de Sousa}}\ \emph {et~al.}(2010)\citenamefont {{de
  Sousa}}, \citenamefont {Steffens}, \citenamefont {Pakter},\ and\
  \citenamefont {Rizzato}}]{deSousa2010}%
  \BibitemOpen
  \bibfield  {author} {\bibinfo {author} {\bibfnamefont {M.~C.}\ \bibnamefont
  {{de Sousa}}}, \bibinfo {author} {\bibfnamefont {F.~M.}\ \bibnamefont
  {Steffens}}, \bibinfo {author} {\bibfnamefont {R.}~\bibnamefont {Pakter}}, \
  and\ \bibinfo {author} {\bibfnamefont {F.~B.}\ \bibnamefont {Rizzato}},\
  }\href {\doibase 10.1103/PhysRevE.82.026402} {\bibfield  {journal} {\bibinfo
  {journal} {Physical Review E}\ }\textbf {\bibinfo {volume} {82}},\ \bibinfo
  {pages} {026402} (\bibinfo {year} {2010})}\BibitemShut {NoStop}%
\bibitem [{\citenamefont {{de Sousa}}\ \emph {et~al.}(2012)\citenamefont {{de
  Sousa}}, \citenamefont {Caldas}, \citenamefont {Rizzato}, \citenamefont
  {Pakter},\ and\ \citenamefont {Steffens}}]{deSousa2012}%
  \BibitemOpen
  \bibfield  {author} {\bibinfo {author} {\bibfnamefont {M.~C.}\ \bibnamefont
  {{de Sousa}}}, \bibinfo {author} {\bibfnamefont {I.~L.}\ \bibnamefont
  {Caldas}}, \bibinfo {author} {\bibfnamefont {F.~B.}\ \bibnamefont {Rizzato}},
  \bibinfo {author} {\bibfnamefont {R.}~\bibnamefont {Pakter}}, \ and\ \bibinfo
  {author} {\bibfnamefont {F.~M.}\ \bibnamefont {Steffens}},\ }\href {\doibase
  10.1103/PhysRevE.86.016217} {\bibfield  {journal} {\bibinfo  {journal}
  {Physical Review E}\ }\textbf {\bibinfo {volume} {86}},\ \bibinfo {pages}
  {016217} (\bibinfo {year} {2012})}\BibitemShut {NoStop}%
\bibitem [{\citenamefont {{de Sousa}}\ and\ \citenamefont
  {Caldas}(2018)}]{deSousa2018}%
  \BibitemOpen
  \bibfield  {author} {\bibinfo {author} {\bibfnamefont {M.~C.}\ \bibnamefont
  {{de Sousa}}}\ and\ \bibinfo {author} {\bibfnamefont {I.~L.}\ \bibnamefont
  {Caldas}},\ }\href {\doibase 10.1063/1.5017508} {\bibfield  {journal}
  {\bibinfo  {journal} {Physics of Plasmas}\ }\textbf {\bibinfo {volume}
  {25}},\ \bibinfo {pages} {043110} (\bibinfo {year} {2018})}\BibitemShut
  {NoStop}%
\bibitem [{\citenamefont {Karney}(1978)}]{Karney1978}%
  \BibitemOpen
  \bibfield  {author} {\bibinfo {author} {\bibfnamefont {C.~F.~F.}\
  \bibnamefont {Karney}},\ }\href {\doibase 10.1063/1.862406} {\bibfield
  {journal} {\bibinfo  {journal} {Physics of Fluids}\ }\textbf {\bibinfo
  {volume} {21}},\ \bibinfo {pages} {1584} (\bibinfo {year}
  {1978})}\BibitemShut {NoStop}%
\bibitem [{\citenamefont {{Corr{\^{e}}a da Silva}}\ \emph
  {et~al.}(2013)\citenamefont {{Corr{\^{e}}a da Silva}}, \citenamefont
  {Pakter}, \citenamefont {Rizzato}, \citenamefont {{de Sousa}}, \citenamefont
  {Caldas},\ and\ \citenamefont {Steffens}}]{CorreaSilva2013}%
  \BibitemOpen
  \bibfield  {author} {\bibinfo {author} {\bibfnamefont {T.~M.}\ \bibnamefont
  {{Corr{\^{e}}a da Silva}}}, \bibinfo {author} {\bibfnamefont
  {R.}~\bibnamefont {Pakter}}, \bibinfo {author} {\bibfnamefont {F.~B.}\
  \bibnamefont {Rizzato}}, \bibinfo {author} {\bibfnamefont {M.~C.}\
  \bibnamefont {{de Sousa}}}, \bibinfo {author} {\bibfnamefont {I.~L.}\
  \bibnamefont {Caldas}}, \ and\ \bibinfo {author} {\bibfnamefont {F.~M.}\
  \bibnamefont {Steffens}},\ }\href {\doibase 10.1103/PhysRevE.88.013101}
  {\bibfield  {journal} {\bibinfo  {journal} {Physical Review E}\ }\textbf
  {\bibinfo {volume} {88}},\ \bibinfo {pages} {013101} (\bibinfo {year}
  {2013})}\BibitemShut {NoStop}%
\bibitem [{\citenamefont {Fisch}(1987)}]{Fisch1987}%
  \BibitemOpen
  \bibfield  {author} {\bibinfo {author} {\bibfnamefont {N.~J.}\ \bibnamefont
  {Fisch}},\ }\href {\doibase 10.1103/RevModPhys.59.175} {\bibfield  {journal}
  {\bibinfo  {journal} {Reviews of Modern Physics}\ }\textbf {\bibinfo {volume}
  {59}},\ \bibinfo {pages} {175} (\bibinfo {year} {1987})}\BibitemShut
  {NoStop}%
\bibitem [{\citenamefont {Berk}, \citenamefont {Breizman},\ and\ \citenamefont
  {Ye}(1992)}]{Berk1992}%
  \BibitemOpen
  \bibfield  {author} {\bibinfo {author} {\bibfnamefont {H.~L.}\ \bibnamefont
  {Berk}}, \bibinfo {author} {\bibfnamefont {B.~N.}\ \bibnamefont {Breizman}},
  \ and\ \bibinfo {author} {\bibfnamefont {H.}~\bibnamefont {Ye}},\ }\href
  {\doibase 10.1103/PhysRevLett.68.3563} {\bibfield  {journal} {\bibinfo
  {journal} {Physical Review Letters}\ }\textbf {\bibinfo {volume} {68}},\
  \bibinfo {pages} {3563} (\bibinfo {year} {1992})}\BibitemShut {NoStop}%
\bibitem [{\citenamefont {Davidson}\ and\ \citenamefont
  {Qin}(2001)}]{Davidson2001}%
  \BibitemOpen
  \bibfield  {author} {\bibinfo {author} {\bibfnamefont {R.~C.}\ \bibnamefont
  {Davidson}}\ and\ \bibinfo {author} {\bibfnamefont {H.}~\bibnamefont {Qin}},\
  }\href@noop {} {\emph {\bibinfo {title} {Physics of intense charged particle
  beams in high energy accelerators}}}\ (\bibinfo  {publisher} {World
  Scientific},\ \bibinfo {address} {London},\ \bibinfo {year}
  {2001})\BibitemShut {NoStop}%
\bibitem [{\citenamefont {Edwards}\ and\ \citenamefont
  {Syphers}(2004)}]{Edwards2004}%
  \BibitemOpen
  \bibfield  {author} {\bibinfo {author} {\bibfnamefont {D.~A.}\ \bibnamefont
  {Edwards}}\ and\ \bibinfo {author} {\bibfnamefont {M.~J.}\ \bibnamefont
  {Syphers}},\ }\href@noop {} {\emph {\bibinfo {title} {An introduction to the
  physics of high energy accelerators}}}\ (\bibinfo  {publisher} {Wiley-VCH},\
  \bibinfo {address} {Weinheim},\ \bibinfo {year} {2004})\BibitemShut {NoStop}%
\bibitem [{\citenamefont {{Gilmour Jr.}}(2011)}]{Gilmour2011}%
  \BibitemOpen
  \bibfield  {author} {\bibinfo {author} {\bibfnamefont {A.~S.}\ \bibnamefont
  {{Gilmour Jr.}}},\ }\href@noop {} {\emph {\bibinfo {title} {Klystrons,
  traveling wave tubes, magnetrons, cross-field amplifiers, and gyrotrons}}}\
  (\bibinfo  {publisher} {Artech House Radar Library},\ \bibinfo {address}
  {Boston},\ \bibinfo {year} {2011})\BibitemShut {NoStop}%
\bibitem [{\citenamefont {Pierce}(1950)}]{Pierce1950}%
  \BibitemOpen
  \bibfield  {author} {\bibinfo {author} {\bibfnamefont {J.~R.}\ \bibnamefont
  {Pierce}},\ }\href@noop {} {\emph {\bibinfo {title} {Traveling Wave Tubes}}}\
  (\bibinfo  {publisher} {Van Nostrand},\ \bibinfo {address} {New York},\
  \bibinfo {year} {1950})\BibitemShut {NoStop}%
\bibitem [{\citenamefont {{Gilmour Jr.}}(1994)}]{Gilmour1994}%
  \BibitemOpen
  \bibfield  {author} {\bibinfo {author} {\bibfnamefont {A.~S.}\ \bibnamefont
  {{Gilmour Jr.}}},\ }\href@noop {} {\emph {\bibinfo {title} {Principles of
  traveling wave tubes}}}\ (\bibinfo  {publisher} {Artech House},\ \bibinfo
  {address} {London},\ \bibinfo {year} {1994})\BibitemShut {NoStop}%
\bibitem [{\citenamefont {Faillon}\ \emph {et~al.}(2008)\citenamefont
  {Faillon}, \citenamefont {Kornfeld}, \citenamefont {Bosch},\ and\
  \citenamefont {Thumm}}]{Faillon2008}%
  \BibitemOpen
  \bibfield  {author} {\bibinfo {author} {\bibfnamefont {G.}~\bibnamefont
  {Faillon}}, \bibinfo {author} {\bibfnamefont {G.}~\bibnamefont {Kornfeld}},
  \bibinfo {author} {\bibfnamefont {E.}~\bibnamefont {Bosch}}, \ and\ \bibinfo
  {author} {\bibfnamefont {M.~K.}\ \bibnamefont {Thumm}},\ }\enquote {\bibinfo
  {title} {Vacuum electronics},}\ \ (\bibinfo  {publisher} {Springer},\
  \bibinfo {address} {Berlin},\ \bibinfo {year} {2008})\ Chap.\ \bibinfo
  {chapter} {Microwave tubes}, pp.\ \bibinfo {pages} {1--82}\BibitemShut
  {NoStop}%
\bibitem [{\citenamefont {Minenna}\ \emph
  {et~al.}(2019{\natexlab{a}})\citenamefont {Minenna}, \citenamefont
  {Andr{\'{e}}}, \citenamefont {Elskens}, \citenamefont {Auboin}, \citenamefont
  {Doveil}, \citenamefont {Puech},\ and\ \citenamefont
  {Duverdier}}]{Minenna2019EPJH}%
  \BibitemOpen
  \bibfield  {author} {\bibinfo {author} {\bibfnamefont {D.~F.~G.}\
  \bibnamefont {Minenna}}, \bibinfo {author} {\bibfnamefont {F.}~\bibnamefont
  {Andr{\'{e}}}}, \bibinfo {author} {\bibfnamefont {Y.}~\bibnamefont
  {Elskens}}, \bibinfo {author} {\bibfnamefont {J.-F.}\ \bibnamefont {Auboin}},
  \bibinfo {author} {\bibfnamefont {F.}~\bibnamefont {Doveil}}, \bibinfo
  {author} {\bibfnamefont {J.}~\bibnamefont {Puech}}, \ and\ \bibinfo {author}
  {\bibfnamefont {{\'{E}}.}~\bibnamefont {Duverdier}},\ }\href {\doibase
  10.1140/epjh/e2018-90023-1} {\bibfield  {journal} {\bibinfo  {journal}
  {European Physical Journal H}\ }\textbf {\bibinfo {volume} {44}},\ \bibinfo
  {pages} {1} (\bibinfo {year} {2019}{\natexlab{a}})}\BibitemShut {NoStop}%
\bibitem [{\citenamefont {Dimonte}\ and\ \citenamefont
  {Malmberg}(1977)}]{Dimonte1977}%
  \BibitemOpen
  \bibfield  {author} {\bibinfo {author} {\bibfnamefont {G.}~\bibnamefont
  {Dimonte}}\ and\ \bibinfo {author} {\bibfnamefont {J.~H.}\ \bibnamefont
  {Malmberg}},\ }\href {\doibase 10.1103/physrevlett.38.401} {\bibfield
  {journal} {\bibinfo  {journal} {Physical Review Letters}\ }\textbf {\bibinfo
  {volume} {38}},\ \bibinfo {pages} {401} (\bibinfo {year} {1977})}\BibitemShut
  {NoStop}%
\bibitem [{\citenamefont {Dimonte}\ and\ \citenamefont
  {Malmberg}(1978)}]{Dimonte1978}%
  \BibitemOpen
  \bibfield  {author} {\bibinfo {author} {\bibfnamefont {G.}~\bibnamefont
  {Dimonte}}\ and\ \bibinfo {author} {\bibfnamefont {J.~H.}\ \bibnamefont
  {Malmberg}},\ }\href {\doibase 10.1063/1.862358} {\bibfield  {journal}
  {\bibinfo  {journal} {Physics of Fluids}\ }\textbf {\bibinfo {volume} {21}},\
  \bibinfo {pages} {1188} (\bibinfo {year} {1978})}\BibitemShut {NoStop}%
\bibitem [{\citenamefont {Tsunoda}, \citenamefont {Doveil},\ and\ \citenamefont
  {Malmberg}(1987)}]{Tsunoda1987}%
  \BibitemOpen
  \bibfield  {author} {\bibinfo {author} {\bibfnamefont {S.~I.}\ \bibnamefont
  {Tsunoda}}, \bibinfo {author} {\bibfnamefont {F.}~\bibnamefont {Doveil}}, \
  and\ \bibinfo {author} {\bibfnamefont {J.~H.}\ \bibnamefont {Malmberg}},\
  }\href {\doibase 10.1103/PhysRevLett.58.1112} {\bibfield  {journal} {\bibinfo
   {journal} {Physical Review Letters}\ }\textbf {\bibinfo {volume} {58}},\
  \bibinfo {pages} {1112} (\bibinfo {year} {1987})}\BibitemShut {NoStop}%
\bibitem [{\citenamefont {Tsunoda}, \citenamefont {Doveil},\ and\ \citenamefont
  {Malmberg}(1991)}]{Tsunoda1991}%
  \BibitemOpen
  \bibfield  {author} {\bibinfo {author} {\bibfnamefont {S.~I.}\ \bibnamefont
  {Tsunoda}}, \bibinfo {author} {\bibfnamefont {F.}~\bibnamefont {Doveil}}, \
  and\ \bibinfo {author} {\bibfnamefont {J.~H.}\ \bibnamefont {Malmberg}},\
  }\href {\doibase 10.1063/1.859911} {\bibfield  {journal} {\bibinfo  {journal}
  {Physics of Fluids B: Plasma Physics}\ }\textbf {\bibinfo {volume} {3}},\
  \bibinfo {pages} {2747} (\bibinfo {year} {1991})}\BibitemShut {NoStop}%
\bibitem [{\citenamefont {Hartmann}\ \emph {et~al.}(1995)\citenamefont
  {Hartmann}, \citenamefont {Driscoll}, \citenamefont {O'Neil},\ and\
  \citenamefont {Shapiro}}]{Hartmann1995}%
  \BibitemOpen
  \bibfield  {author} {\bibinfo {author} {\bibfnamefont {D.~A.}\ \bibnamefont
  {Hartmann}}, \bibinfo {author} {\bibfnamefont {C.~F.}\ \bibnamefont
  {Driscoll}}, \bibinfo {author} {\bibfnamefont {T.~M.}\ \bibnamefont
  {O'Neil}}, \ and\ \bibinfo {author} {\bibfnamefont {V.~D.}\ \bibnamefont
  {Shapiro}},\ }\href {\doibase 10.1063/1.871418} {\bibfield  {journal}
  {\bibinfo  {journal} {Physics of Plasmas}\ }\textbf {\bibinfo {volume} {2}},\
  \bibinfo {pages} {654} (\bibinfo {year} {1995})}\BibitemShut {NoStop}%
\bibitem [{\citenamefont {Guyomarc’h}(1996)}]{Guyomarch1996}%
  \BibitemOpen
  \bibfield  {author} {\bibinfo {author} {\bibfnamefont {D.}~\bibnamefont
  {Guyomarc’h}},\ }\emph {\bibinfo {title} {Un tube {\`{a}} onde progressive
  pour l’{\'{e}}tude de la turbulence plasma}},\ \href@noop {} {Ph.D.
  thesis},\ \bibinfo  {school} {Universit{\'{e}} de Provence}, \bibinfo
  {address} {Marseilles, France} (\bibinfo {year} {1996})\BibitemShut {NoStop}%
\bibitem [{\citenamefont {Doveil}, \citenamefont {Escande},\ and\ \citenamefont
  {Macor}(2005)}]{Doveil2005PRL}%
  \BibitemOpen
  \bibfield  {author} {\bibinfo {author} {\bibfnamefont {F.}~\bibnamefont
  {Doveil}}, \bibinfo {author} {\bibfnamefont {D.~F.}\ \bibnamefont {Escande}},
  \ and\ \bibinfo {author} {\bibfnamefont {A.}~\bibnamefont {Macor}},\ }\href
  {\doibase 10.1103/physrevlett.94.085003} {\bibfield  {journal} {\bibinfo
  {journal} {Physical Review Letters}\ }\textbf {\bibinfo {volume} {94}},\
  \bibinfo {pages} {085003} (\bibinfo {year} {2005})}\BibitemShut {NoStop}%
\bibitem [{\citenamefont {Doveil}, \citenamefont {Macor},\ and\ \citenamefont
  {Auhmani}(2005)}]{Doveil2005PPCF}%
  \BibitemOpen
  \bibfield  {author} {\bibinfo {author} {\bibfnamefont {F.}~\bibnamefont
  {Doveil}}, \bibinfo {author} {\bibfnamefont {A.}~\bibnamefont {Macor}}, \
  and\ \bibinfo {author} {\bibfnamefont {K.}~\bibnamefont {Auhmani}},\ }\href
  {\doibase 10.1088/0741-3335/47/5a/018} {\bibfield  {journal} {\bibinfo
  {journal} {Plasma Physics and Controlled Fusion}\ }\textbf {\bibinfo {volume}
  {47}},\ \bibinfo {pages} {A261} (\bibinfo {year} {2005})}\BibitemShut
  {NoStop}%
\bibitem [{\citenamefont {Macor}(2007)}]{MacorThesis2007}%
  \BibitemOpen
  \bibfield  {author} {\bibinfo {author} {\bibfnamefont {A.}~\bibnamefont
  {Macor}},\ }\emph {\bibinfo {title} {D'un faisceau test {\`{a}}
  l'auto-coh{\'{e}}rence dans l'interaction onde-particule}},\ \href@noop {}
  {Ph.D. thesis},\ \bibinfo  {school} {Universit{\'{e}} de Provence}, \bibinfo
  {address} {Marseilles, France} (\bibinfo {year} {2007})\BibitemShut {NoStop}%
\bibitem [{\citenamefont {Doveil}\ and\ \citenamefont
  {Macor}(2011)}]{Doveil2011}%
  \BibitemOpen
  \bibfield  {author} {\bibinfo {author} {\bibfnamefont {F.}~\bibnamefont
  {Doveil}}\ and\ \bibinfo {author} {\bibfnamefont {A.}~\bibnamefont {Macor}},\
  }\href {\doibase 10.1103/PhysRevE.84.045401} {\bibfield  {journal} {\bibinfo
  {journal} {Physical Review E}\ }\textbf {\bibinfo {volume} {84}},\ \bibinfo
  {pages} {045401} (\bibinfo {year} {2011})}\BibitemShut {NoStop}%
\bibitem [{\citenamefont {Nordsieck}(1953)}]{Nordsieck1953}%
  \BibitemOpen
  \bibfield  {author} {\bibinfo {author} {\bibfnamefont {A.}~\bibnamefont
  {Nordsieck}},\ }\href {\doibase 10.1109/jrproc.1953.274404} {\bibfield
  {journal} {\bibinfo  {journal} {Proceedings of the {IRE}}\ }\textbf {\bibinfo
  {volume} {41}},\ \bibinfo {pages} {630} (\bibinfo {year} {1953})}\BibitemShut
  {NoStop}%
\bibitem [{\citenamefont {Tien}(1956)}]{Tien1956}%
  \BibitemOpen
  \bibfield  {author} {\bibinfo {author} {\bibfnamefont {P.~K.}\ \bibnamefont
  {Tien}},\ }\href {\doibase 10.1002/j.1538-7305.1956.tb02386.x} {\bibfield
  {journal} {\bibinfo  {journal} {Bell System Technical Journal}\ }\textbf
  {\bibinfo {volume} {35}},\ \bibinfo {pages} {349} (\bibinfo {year}
  {1956})}\BibitemShut {NoStop}%
\bibitem [{\citenamefont {{O'Neil}}, \citenamefont {Winfrey},\ and\
  \citenamefont {Malmberg}(1971)}]{ONeil1971}%
  \BibitemOpen
  \bibfield  {author} {\bibinfo {author} {\bibfnamefont {T.~M.}\ \bibnamefont
  {{O'Neil}}}, \bibinfo {author} {\bibfnamefont {J.~H.}\ \bibnamefont
  {Winfrey}}, \ and\ \bibinfo {author} {\bibfnamefont {J.~H.}\ \bibnamefont
  {Malmberg}},\ }\href {\doibase 10.1063/1.1693587} {\bibfield  {journal}
  {\bibinfo  {journal} {Physics of Fluids}\ }\textbf {\bibinfo {volume} {14}},\
  \bibinfo {pages} {1204} (\bibinfo {year} {1971})}\BibitemShut {NoStop}%
\bibitem [{\citenamefont {{O'Neil}}\ and\ \citenamefont
  {Winfrey}(1972)}]{ONeil1972}%
  \BibitemOpen
  \bibfield  {author} {\bibinfo {author} {\bibfnamefont {T.~M.}\ \bibnamefont
  {{O'Neil}}}\ and\ \bibinfo {author} {\bibfnamefont {J.~H.}\ \bibnamefont
  {Winfrey}},\ }\href {\doibase 10.1063/1.1694117} {\bibfield  {journal}
  {\bibinfo  {journal} {Physics of Fluids}\ }\textbf {\bibinfo {volume} {15}},\
  \bibinfo {pages} {1514} (\bibinfo {year} {1972})}\BibitemShut {NoStop}%
\bibitem [{\citenamefont {Chandre}\ \emph {et~al.}(2005)\citenamefont
  {Chandre}, \citenamefont {Ciraolo}, \citenamefont {Doveil}, \citenamefont
  {Lima}, \citenamefont {Macor},\ and\ \citenamefont {Vittot}}]{Chandre2005}%
  \BibitemOpen
  \bibfield  {author} {\bibinfo {author} {\bibfnamefont {C.}~\bibnamefont
  {Chandre}}, \bibinfo {author} {\bibfnamefont {G.}~\bibnamefont {Ciraolo}},
  \bibinfo {author} {\bibfnamefont {F.}~\bibnamefont {Doveil}}, \bibinfo
  {author} {\bibfnamefont {R.}~\bibnamefont {Lima}}, \bibinfo {author}
  {\bibfnamefont {A.}~\bibnamefont {Macor}}, \ and\ \bibinfo {author}
  {\bibfnamefont {M.}~\bibnamefont {Vittot}},\ }\href {\doibase
  10.1103/PhysRevLett.94.074101} {\bibfield  {journal} {\bibinfo  {journal}
  {Physical Review Letters}\ }\textbf {\bibinfo {volume} {94}},\ \bibinfo
  {pages} {074101} (\bibinfo {year} {2005})}\BibitemShut {NoStop}%
\bibitem [{\citenamefont {Doveil}, \citenamefont {Macor},\ and\ \citenamefont
  {Elskens}(2006)}]{Doveil2006}%
  \BibitemOpen
  \bibfield  {author} {\bibinfo {author} {\bibfnamefont {F.}~\bibnamefont
  {Doveil}}, \bibinfo {author} {\bibfnamefont {A.}~\bibnamefont {Macor}}, \
  and\ \bibinfo {author} {\bibfnamefont {Y.}~\bibnamefont {Elskens}},\ }\href
  {\doibase 10.1063/1.2216850} {\bibfield  {journal} {\bibinfo  {journal}
  {Chaos}\ }\textbf {\bibinfo {volume} {16}},\ \bibinfo {pages} {033103}
  (\bibinfo {year} {2006})}\BibitemShut {NoStop}%
\bibitem [{\citenamefont {Macor}\ \emph {et~al.}(2007)\citenamefont {Macor},
  \citenamefont {Doveil}, \citenamefont {Chandre}, \citenamefont {Ciraolo},
  \citenamefont {Lima},\ and\ \citenamefont {Vittot}}]{MacorEPJD2007}%
  \BibitemOpen
  \bibfield  {author} {\bibinfo {author} {\bibfnamefont {A.}~\bibnamefont
  {Macor}}, \bibinfo {author} {\bibfnamefont {F.}~\bibnamefont {Doveil}},
  \bibinfo {author} {\bibfnamefont {C.}~\bibnamefont {Chandre}}, \bibinfo
  {author} {\bibfnamefont {G.}~\bibnamefont {Ciraolo}}, \bibinfo {author}
  {\bibfnamefont {R.}~\bibnamefont {Lima}}, \ and\ \bibinfo {author}
  {\bibfnamefont {M.}~\bibnamefont {Vittot}},\ }\href {\doibase
  10.1140/epjd/e2006-00260-6} {\bibfield  {journal} {\bibinfo  {journal} {The
  European Physical Journal D}\ }\textbf {\bibinfo {volume} {41}},\ \bibinfo
  {pages} {519} (\bibinfo {year} {2007})}\BibitemShut {NoStop}%
\bibitem [{\citenamefont {Macor}, \citenamefont {Doveil},\ and\ \citenamefont
  {Garabedian}(2007)}]{MacorNPCS2007}%
  \BibitemOpen
  \bibfield  {author} {\bibinfo {author} {\bibfnamefont {A.}~\bibnamefont
  {Macor}}, \bibinfo {author} {\bibfnamefont {F.}~\bibnamefont {Doveil}}, \
  and\ \bibinfo {author} {\bibfnamefont {E.}~\bibnamefont {Garabedian}},\
  }\href@noop {} {\bibfield  {journal} {\bibinfo  {journal} {Nonlinear
  Phenomena in Complex Systems}\ }\textbf {\bibinfo {volume} {10}},\ \bibinfo
  {pages} {180} (\bibinfo {year} {2007})}\BibitemShut {NoStop}%
\bibitem [{\citenamefont {Tennyson}, \citenamefont {Meiss},\ and\ \citenamefont
  {Morrison}(1994)}]{Tennyson1994}%
  \BibitemOpen
  \bibfield  {author} {\bibinfo {author} {\bibfnamefont {J.~L.}\ \bibnamefont
  {Tennyson}}, \bibinfo {author} {\bibfnamefont {J.~D.}\ \bibnamefont {Meiss}},
  \ and\ \bibinfo {author} {\bibfnamefont {P.~J.}\ \bibnamefont {Morrison}},\
  }\href {\doibase 10.1016/0167-2789(94)90178-3} {\bibfield  {journal}
  {\bibinfo  {journal} {Physica D: Nonlinear Phenomena}\ }\textbf {\bibinfo
  {volume} {71}},\ \bibinfo {pages} {1} (\bibinfo {year} {1994})}\BibitemShut
  {NoStop}%
\bibitem [{\citenamefont {del Castillo-Negrete}\ and\ \citenamefont
  {Firpo}(2002)}]{del-Castillo2002}%
  \BibitemOpen
  \bibfield  {author} {\bibinfo {author} {\bibfnamefont {D.}~\bibnamefont {del
  Castillo-Negrete}}\ and\ \bibinfo {author} {\bibfnamefont {M.-C.}\
  \bibnamefont {Firpo}},\ }\href {\doibase 10.1063/1.1470203} {\bibfield
  {journal} {\bibinfo  {journal} {Chaos}\ }\textbf {\bibinfo {volume} {12}},\
  \bibinfo {pages} {496} (\bibinfo {year} {2002})}\BibitemShut {NoStop}%
\bibitem [{\citenamefont {Vedenov}, \citenamefont {Velikhov},\ and\
  \citenamefont {Sagdeev}(1962)}]{Vedenov1962}%
  \BibitemOpen
  \bibfield  {author} {\bibinfo {author} {\bibfnamefont {A.~A.}\ \bibnamefont
  {Vedenov}}, \bibinfo {author} {\bibfnamefont {E.~P.}\ \bibnamefont
  {Velikhov}}, \ and\ \bibinfo {author} {\bibfnamefont {R.~Z.}\ \bibnamefont
  {Sagdeev}},\ }\href@noop {} {\bibfield  {journal} {\bibinfo  {journal}
  {Nuclear Fusion Supplement}\ }\textbf {\bibinfo {volume} {2}},\ \bibinfo
  {pages} {465} (\bibinfo {year} {1962})}\BibitemShut {NoStop}%
\bibitem [{\citenamefont {Drummond}\ and\ \citenamefont
  {Pines}(1962)}]{Drummond1962}%
  \BibitemOpen
  \bibfield  {author} {\bibinfo {author} {\bibfnamefont {W.~E.}\ \bibnamefont
  {Drummond}}\ and\ \bibinfo {author} {\bibfnamefont {D.}~\bibnamefont
  {Pines}},\ }\href@noop {} {\bibfield  {journal} {\bibinfo  {journal} {Nuclear
  Fusion Supplement}\ }\textbf {\bibinfo {volume} {3}},\ \bibinfo {pages}
  {1049} (\bibinfo {year} {1962})}\BibitemShut {NoStop}%
\bibitem [{\citenamefont {Vedenov}(1963)}]{Vedenov1963}%
  \BibitemOpen
  \bibfield  {author} {\bibinfo {author} {\bibfnamefont {A.~A.}\ \bibnamefont
  {Vedenov}},\ }\href {\doibase 10.1088/0368-3281/5/3/305} {\bibfield
  {journal} {\bibinfo  {journal} {Journal of Nuclear Energy, Part C Plasma
  Physics}\ }\textbf {\bibinfo {volume} {5}},\ \bibinfo {pages} {169} (\bibinfo
  {year} {1963})}\BibitemShut {NoStop}%
\bibitem [{\citenamefont {Elskens}(2007)}]{Elskens2007}%
  \BibitemOpen
  \bibfield  {author} {\bibinfo {author} {\bibfnamefont {Y.}~\bibnamefont
  {Elskens}},\ }\href@noop {} {\bibfield  {journal} {\bibinfo  {journal}
  {Physics AUC}\ }\textbf {\bibinfo {volume} {17}},\ \bibinfo {pages} {109}
  (\bibinfo {year} {2007})}\BibitemShut {NoStop}%
\bibitem [{\citenamefont {Elskens}(2010)}]{Elskens2010}%
  \BibitemOpen
  \bibfield  {author} {\bibinfo {author} {\bibfnamefont {Y.}~\bibnamefont
  {Elskens}},\ }\href {\doibase 10.1016/j.cnsns.2008.05.014} {\bibfield
  {journal} {\bibinfo  {journal} {Communications in Nonlinear Science and
  Numerical Simulation}\ }\textbf {\bibinfo {volume} {15}},\ \bibinfo {pages}
  {10} (\bibinfo {year} {2010})}\BibitemShut {NoStop}%
\bibitem [{\citenamefont {Elskens}\ and\ \citenamefont
  {Pardoux}(2010)}]{Elskens2010AAP}%
  \BibitemOpen
  \bibfield  {author} {\bibinfo {author} {\bibfnamefont {Y.}~\bibnamefont
  {Elskens}}\ and\ \bibinfo {author} {\bibfnamefont {E.}~\bibnamefont
  {Pardoux}},\ }\href {\doibase 10.1214/09-AAP671} {\bibfield  {journal}
  {\bibinfo  {journal} {Annals of Applied Probability}\ }\textbf {\bibinfo
  {volume} {20}},\ \bibinfo {pages} {2022} (\bibinfo {year}
  {2010})}\BibitemShut {NoStop}%
\bibitem [{\citenamefont {Besse}\ \emph {et~al.}(2011)\citenamefont {Besse},
  \citenamefont {Elskens}, \citenamefont {Escande},\ and\ \citenamefont
  {Bertrand}}]{Besse2011}%
  \BibitemOpen
  \bibfield  {author} {\bibinfo {author} {\bibfnamefont {N.}~\bibnamefont
  {Besse}}, \bibinfo {author} {\bibfnamefont {Y.}~\bibnamefont {Elskens}},
  \bibinfo {author} {\bibfnamefont {D.~F.}\ \bibnamefont {Escande}}, \ and\
  \bibinfo {author} {\bibfnamefont {P.}~\bibnamefont {Bertrand}},\ }\href
  {\doibase 10.1088/0741-3335/53/2/025012} {\bibfield  {journal} {\bibinfo
  {journal} {Plasma Physics and Controlled Fusion}\ }\textbf {\bibinfo {volume}
  {53}},\ \bibinfo {pages} {025012} (\bibinfo {year} {2011})}\BibitemShut
  {NoStop}%
\bibitem [{\citenamefont {Elskens}(2012)}]{Elskens2012}%
  \BibitemOpen
  \bibfield  {author} {\bibinfo {author} {\bibfnamefont {Y.}~\bibnamefont
  {Elskens}},\ }\href {\doibase 10.1007/s10955-012-0546-2} {\bibfield
  {journal} {\bibinfo  {journal} {Journal of Statistical Physics}\ }\textbf
  {\bibinfo {volume} {148}},\ \bibinfo {pages} {591} (\bibinfo {year}
  {2012})}\BibitemShut {NoStop}%
\bibitem [{\citenamefont {Andr{\'{e}}}\ \emph {et~al.}(2013)\citenamefont
  {Andr{\'{e}}}, \citenamefont {Bernardi}, \citenamefont {Ryskin},
  \citenamefont {Doveil},\ and\ \citenamefont {Elskens}}]{Andre2013}%
  \BibitemOpen
  \bibfield  {author} {\bibinfo {author} {\bibfnamefont {F.}~\bibnamefont
  {Andr{\'{e}}}}, \bibinfo {author} {\bibfnamefont {P.}~\bibnamefont
  {Bernardi}}, \bibinfo {author} {\bibfnamefont {N.~M.}\ \bibnamefont
  {Ryskin}}, \bibinfo {author} {\bibfnamefont {F.}~\bibnamefont {Doveil}}, \
  and\ \bibinfo {author} {\bibfnamefont {Y.}~\bibnamefont {Elskens}},\ }\href
  {\doibase 10.1209/0295-5075/103/28004} {\bibfield  {journal} {\bibinfo
  {journal} {Europhysics Letters}\ }\textbf {\bibinfo {volume} {103}},\
  \bibinfo {pages} {28004} (\bibinfo {year} {2013})}\BibitemShut {NoStop}%
\bibitem [{\citenamefont {Minenna}\ \emph {et~al.}(2018)\citenamefont
  {Minenna}, \citenamefont {Elskens}, \citenamefont {Andr{\'{e}}},\ and\
  \citenamefont {Doveil}}]{Minenna2018}%
  \BibitemOpen
  \bibfield  {author} {\bibinfo {author} {\bibfnamefont {D.~F.~G.}\
  \bibnamefont {Minenna}}, \bibinfo {author} {\bibfnamefont {Y.}~\bibnamefont
  {Elskens}}, \bibinfo {author} {\bibfnamefont {F.}~\bibnamefont
  {Andr{\'{e}}}}, \ and\ \bibinfo {author} {\bibfnamefont {F.}~\bibnamefont
  {Doveil}},\ }\href {\doibase 10.1209/0295-5075/122/44002} {\bibfield
  {journal} {\bibinfo  {journal} {Europhysics Letters}\ }\textbf {\bibinfo
  {volume} {122}},\ \bibinfo {pages} {44002} (\bibinfo {year}
  {2018})}\BibitemShut {NoStop}%
\bibitem [{\citenamefont {Minenna}\ \emph
  {et~al.}(2019{\natexlab{b}})\citenamefont {Minenna}, \citenamefont {Elskens},
  \citenamefont {Andr{\'{e}}}, \citenamefont {Poy{\'{e}}}, \citenamefont
  {Puech},\ and\ \citenamefont {Doveil}}]{Minenna2019IEEE}%
  \BibitemOpen
  \bibfield  {author} {\bibinfo {author} {\bibfnamefont {D.~F.~G.}\
  \bibnamefont {Minenna}}, \bibinfo {author} {\bibfnamefont {Y.}~\bibnamefont
  {Elskens}}, \bibinfo {author} {\bibfnamefont {F.}~\bibnamefont
  {Andr{\'{e}}}}, \bibinfo {author} {\bibfnamefont {A.}~\bibnamefont
  {Poy{\'{e}}}}, \bibinfo {author} {\bibfnamefont {J.}~\bibnamefont {Puech}}, \
  and\ \bibinfo {author} {\bibfnamefont {F.}~\bibnamefont {Doveil}},\ }\href
  {\doibase 10.1109/TED.2019.2928450} {\bibfield  {journal} {\bibinfo
  {journal} {IEEE Transactions on Electron Devices}\ }\textbf {\bibinfo
  {volume} {66}},\ \bibinfo {pages} {4042} (\bibinfo {year}
  {2019}{\natexlab{b}})}\BibitemShut {NoStop}%
\bibitem [{\citenamefont {{Guyomarc’h}}\ and\ \citenamefont
  {Doveil}(2000)}]{Guyomarch2000}%
  \BibitemOpen
  \bibfield  {author} {\bibinfo {author} {\bibfnamefont {D.}~\bibnamefont
  {{Guyomarc’h}}}\ and\ \bibinfo {author} {\bibfnamefont {F.}~\bibnamefont
  {Doveil}},\ }\href {\doibase 10.1063/1.1319339} {\bibfield  {journal}
  {\bibinfo  {journal} {Review of Scientific Instruments}\ }\textbf {\bibinfo
  {volume} {71}},\ \bibinfo {pages} {4087} (\bibinfo {year}
  {2000})}\BibitemShut {NoStop}%
\bibitem [{\citenamefont {Spohn}(2004)}]{Spohn2004}%
  \BibitemOpen
  \bibfield  {author} {\bibinfo {author} {\bibfnamefont {H.}~\bibnamefont
  {Spohn}},\ }\href@noop {} {\emph {\bibinfo {title} {Dynamics of charged
  particles and their radiation fields}}}\ (\bibinfo  {publisher} {Cambridge
  University Press},\ \bibinfo {address} {Cambridge},\ \bibinfo {year}
  {2004})\BibitemShut {NoStop}%
\bibitem [{\citenamefont {Jackson}(1999)}]{Jackson1999}%
  \BibitemOpen
  \bibfield  {author} {\bibinfo {author} {\bibfnamefont {J.~D.}\ \bibnamefont
  {Jackson}},\ }\href@noop {} {\emph {\bibinfo {title} {Classical
  electrodynamics}}},\ \bibinfo {edition} {3rd}\ ed.\ (\bibinfo  {publisher}
  {Wiley},\ \bibinfo {address} {New York},\ \bibinfo {year} {1999})\BibitemShut
  {NoStop}%
\bibitem [{\citenamefont {Malmberg}\ and\ \citenamefont
  {Wharton}(1969)}]{Malmberg1969}%
  \BibitemOpen
  \bibfield  {author} {\bibinfo {author} {\bibfnamefont {J.~H.}\ \bibnamefont
  {Malmberg}}\ and\ \bibinfo {author} {\bibfnamefont {C.~B.}\ \bibnamefont
  {Wharton}},\ }\href {\doibase 10.1063/1.1692402} {\bibfield  {journal}
  {\bibinfo  {journal} {Physics of Fluids}\ }\textbf {\bibinfo {volume} {12}},\
  \bibinfo {pages} {2600} (\bibinfo {year} {1969})}\BibitemShut {NoStop}%
\bibitem [{\citenamefont {Kompfner}(1950)}]{Kompfner1950}%
  \BibitemOpen
  \bibfield  {author} {\bibinfo {author} {\bibfnamefont {R.}~\bibnamefont
  {Kompfner}},\ }\href {\doibase 10.1049/jbire.1950.0028} {\bibfield  {journal}
  {\bibinfo  {journal} {Journal of the British Institution of Radio Engineers}\
  }\textbf {\bibinfo {volume} {10}},\ \bibinfo {pages} {283} (\bibinfo {year}
  {1950})}\BibitemShut {NoStop}%
\bibitem [{\citenamefont {Johnson}(1955)}]{Johnson1955}%
  \BibitemOpen
  \bibfield  {author} {\bibinfo {author} {\bibfnamefont {H.~R.}\ \bibnamefont
  {Johnson}},\ }\href {\doibase 10.1109/JRPROC.1955.278155} {\bibfield
  {journal} {\bibinfo  {journal} {Proceedings of the IRE}\ }\textbf {\bibinfo
  {volume} {43}},\ \bibinfo {pages} {874} (\bibinfo {year} {1955})}\BibitemShut
  {NoStop}%
\bibitem [{\citenamefont {Birdsall}\ and\ \citenamefont
  {Brewer}(1954)}]{Birdsall1954}%
  \BibitemOpen
  \bibfield  {author} {\bibinfo {author} {\bibfnamefont {C.~K.}\ \bibnamefont
  {Birdsall}}\ and\ \bibinfo {author} {\bibfnamefont {G.~R.}\ \bibnamefont
  {Brewer}},\ }\href {\doibase 10.1109/T-ED.1954.14017} {\bibfield  {journal}
  {\bibinfo  {journal} {IRE Transactions on Electron Devices}\ }\textbf
  {\bibinfo {volume} {ED-1}},\ \bibinfo {pages} {1} (\bibinfo {year}
  {1954})}\BibitemShut {NoStop}%
\bibitem [{\citenamefont {Branch}\ and\ \citenamefont
  {Mihran}(1955)}]{Branch1955}%
  \BibitemOpen
  \bibfield  {author} {\bibinfo {author} {\bibfnamefont {G.~M.}\ \bibnamefont
  {Branch}}\ and\ \bibinfo {author} {\bibfnamefont {T.~G.}\ \bibnamefont
  {Mihran}},\ }\href {\doibase 10.1109/T-ED.1955.14065} {\bibfield  {journal}
  {\bibinfo  {journal} {IRE Transactions on Electron Devices}\ }\textbf
  {\bibinfo {volume} {ED-2}},\ \bibinfo {pages} {3} (\bibinfo {year}
  {1955})}\BibitemShut {NoStop}%
\bibitem [{\citenamefont {Minenna}\ \emph
  {et~al.}(2019{\natexlab{c}})\citenamefont {Minenna}, \citenamefont
  {Terentyuk}, \citenamefont {Andr{\'e}}, \citenamefont {Elskens},\ and\
  \citenamefont {Ryskin}}]{Minenna2019PhysScr}%
  \BibitemOpen
  \bibfield  {author} {\bibinfo {author} {\bibfnamefont {D.~F.~G.}\
  \bibnamefont {Minenna}}, \bibinfo {author} {\bibfnamefont {A.~G.}\
  \bibnamefont {Terentyuk}}, \bibinfo {author} {\bibfnamefont {F.}~\bibnamefont
  {Andr{\'e}}}, \bibinfo {author} {\bibfnamefont {Y.}~\bibnamefont {Elskens}},
  \ and\ \bibinfo {author} {\bibfnamefont {N.~M.}\ \bibnamefont {Ryskin}},\
  }\href {\doibase 10.1088/1402-4896/ab060e} {\bibfield  {journal} {\bibinfo
  {journal} {Physica Scripta}\ }\textbf {\bibinfo {volume} {94}},\ \bibinfo
  {pages} {055601} (\bibinfo {year} {2019}{\natexlab{c}})}\BibitemShut
  {NoStop}%
\end{thebibliography}%

\end{document}